\documentclass[journal,comsoc]{IEEEtran}
\usepackage[T1]{fontenc}
\usepackage {multirow}
\usepackage [english]{babel}
\usepackage{fixltx2e}
\usepackage{setspace}
\usepackage{booktabs}
\usepackage{array}
\usepackage{tabularx}
\usepackage{calc}
\usepackage{makecell}
\usepackage{upgreek}

\usepackage{xcolor}
\usepackage{multirow}
\usepackage{booktabs}
\usepackage{array}
\newcolumntype{L}[1]{>{\raggedright\let\newline\\\arraybackslash\hspace{0pt}}m{#1}}
\newcolumntype{C}[1]{>{\centering\let\newline\\\arraybackslash\hspace{0pt}}m{#1}}
\newcolumntype{R}[1]{>{\raggedleft\let\newline\\\arraybackslash\hspace{0pt}}m{#1}}

\usepackage{cite}

\ifCLASSINFOpdf
   \usepackage[pdftex]{graphicx}
\else
   or other class option (dvipsone, dvipdf, if not using dvips). graphicx
   \usepackage[dvips]{graphicx}
   \graphicspath{{../eps/}}
   \DeclareGraphicsExtensions{.eps}
\fi
\usepackage{amsmath}
\usepackage[cmintegrals]{newtxmath}
\usepackage{algorithmic}

\usepackage{array}
\usepackage{makecell}
\usepackage{verbatim}

\ifCLASSOPTIONcompsoc
  \usepackage[caption=false,font=normalsize,labelfont=sf,textfont=sf]{subfig}
\else
  \usepackage[caption=false,font=footnotesize]{subfig}
\fi
\begin{document}
\setcounter{figure}{0}
\renewcommand{\figurename}{Fig.}
\renewcommand{\thefigure}{\arabic{figure}}

\title{Sub-Nyquist Sampling OFDM Radar With a Time-Frequency Phase-Coded Waveform}
%
%

\author{Seonghyeon~Kang,~\IEEEmembership{Student Member,~IEEE,}
        Kawon~Han,~\IEEEmembership{Member,~IEEE,}
        and~Songcheol~Hong,~\IEEEmembership{Member,~IEEE}

\thanks{Manuscript received xx, 2024. This work is supported in part by the Institute of Information \& Communications Technology Planning and Evaluation (IITP) funded by the Korea Government, Ministry of Science and ICT (MSIT) under Grant 2019-0-00826. \emph{(Corresponding author: Kawon Han.)}

Seonghyeon Kang and Songcheol Hong are with the Department of Electrical Engineering, Korea Advanced Institute of Science and Technology (KAIST), Daejeon 34141, South Korea (e-mail: stx1226@kaist.ac.kr;songcheol1234@gmail.com).

Kawon Han is with the Department of Electronics and Electrical Engineering, University College London (UCL), London, United Kingdom (e-mail: hkw1176@gmail.com).}}

\maketitle

\begin{abstract}
This paper presents a time-frequency phase-coded sub-Nyquist sampling orthogonal frequency division multiplexing (PC-SNS-OFDM) radar system to reduce the analog-to-digital converter (ADC) sampling rate without any additional hardware or signal processing. The proposed radar divides the transmitted OFDM signal into multiple sub-bands along the frequency axis and provides orthogonality to these sub-bands by multiplying phase codes in both the time and frequency domains. Although the sampling rate is reduced by the factor of the number of sub-bands, the sub-bands above the sampling rate are folded into the lowest one due to aliasing. In the process of restoring the signals in folded sub-bands to those in full signal bands, the proposed PC-SNS-OFDM radar effectively eliminates symbol-mismatch noise while introducing trade-offs in the range and Doppler ambiguities. The utilization of phase codes in both the frequency and time domains provides flexible control of the range and Doppler ambiguities. It also improves the signal-to-noise ratio (SNR) of detected targets compared to an earlier sub-Nyquist sampling OFDM radar system. This is validated with simulations and experiments under various sub-Nyquist sampling rates.
\end{abstract}


%
\IEEEpeerreviewmaketitle

\section{Introduction}
%
%
%
%
\IEEEPARstart{O}RTHOGONAL frequency division multiplexing (OFDM) radar has been attracting keen attention as a future digital radar system. It is known that OFDM radar utilizing multiple orthogonal subcarriers offers considerable flexibility in terms of signal generation and processing \cite{flexible1, flexible2, flexible3, flexible4, flexible5, flexible6, MIMO1, MIMO2}. Phase-modulated continuous wave (PMCW) radar has also been studied as another digital radar candidate \cite{PMCW1, PMCW2, PMCW3}. \textcolor{black}{OFDM radar can utilize advanced modulation schemes \textcolor{black}{while PMCW uses} binary phase modulation. Moreover, OFDM radar signals are well band limited and occupy a smaller frequency bandwidth than those of PMCW \cite{OFDM1}.} Furthermore, its capability for joint communication and sensing (JCAS) is an unmatched feature which is expected to play a crucial role in 6G communication and autonomous driving \cite{Radcom1, Radcom2, Radcom3, Radcom4, Radcom5, Radcom6, Radcom7, Radcom8, Radcom9, Radcom10}. 

Despite these advantages, there is a significant challenge that must be addressed. This can be explained through performance measures of OFDM radar systems \cite{Parameter, Parameter2}. Radar performance metrics include the range resolution, maximum detectable range, velocity resolution, and maximum detectable velocity. Note that a wide bandwidth signal is necessary to attain a high range resolution. The analog-to-digital converters (ADCs) used in OFDM radar systems must have a sampling rate that exceeds the signal bandwidth. This differs from analog radar, of which the sampling rates depend on the maximum detectable range. It is expected that a sampling rate of \textcolor{black}{$2-4$}  GHz will be required \cite{Digital} to ensure cm-level range resolutions for OFDM radar. Moreover, imaging radar systems must have large antenna apertures to provide high angular resolutions, which are implemented with multiple channels of a multi-input multi-output (MIMO) radar system. The number of ADCs must equal that of the channels, which complicates the implementation and increases the amount of processing data \cite{Angle1, Angle2}.

To reduce the ADC sampling rate in an OFDM radar system, several approaches have been reported \cite{SC_OFDM1, SC_OFDM2, SC_OFDM3, SC_OFDM4, SC_OFDM5, SC_OFDM6, SC_OFDM7, sparse_OFDM, FC_OFDM, FC_OFDM2, SA_OFDM, SNS_OFDM}. One strategy uses stepped-carrier (SC) OFDM  radar, as described in the literature \cite{SC_OFDM1, SC_OFDM2, SC_OFDM3, SC_OFDM4, SC_OFDM5, SC_OFDM6, SC_OFDM7}. This approach utilizes the time-interleaving transmission of a wide-bandwidth OFDM signal by dividing it into several narrow sub-bands. However, the sequential transmission of sub-band signals results in a decrease in the maximum unambiguous velocity. Furthermore, a signal source with a fast-settling-time phase-lock loop (PLL) is required to take advantage of all subcarriers in the receiver \cite{SC_OFDM5, SC_OFDM6}. Also, phase-offset calibration between successive sub-bands is necessary to ensure the continuity of phase variations along the subcarriers.

A sparse OFDM radar system is also introduced as an approach to reduce the baseband bandwidth of the OFDM radar signal \cite{sparse_OFDM}. Sparse OFDM radar employs a baseband signal comprising narrow-band OFDM signals with different bandwidths centered at randomly chosen discrete frequencies. The baseband signal is up-converted to the local oscillator (LO) hopping frequencies, resulting in a sparse OFDM signal. The sparse OFDM signal is reconstructed through compressed sensing (CS) to achieve a resolution equivalent to that of full OFDM radar. However, LO frequency hopping leads to increased hardware complexity, while implementing CS demands significant computational effort.

Another approach with a reduced baseband sampling rate is to use frequency-comb (FC) OFDM  radar, as discussed in \cite{FC_OFDM, FC_OFDM2}. This system increases the RF bandwidth by multiplying the baseband OFDM signal by a frequency comb consisting of $L$ multiple carriers, of which the spacing is equivalent to the bandwidth of the baseband signal. However, this method necessitates complex up- and down-conversion hardware due to the generation of the comb carrier frequency. The wide bandwidth must be sampled without increasing the sampling rate of the ADC. Therefore, only every \textcolor{black}{$L\text{th}$} subcarrier of the transmitter (Tx) baseband signal is utilized, leading to a reduction of the maximum unambiguous range and range processing gain. Furthermore, this approach introduces a phase offset between different comb frequencies, which requires a precise calibration process. Inaccurate calibration results in significant degradation of the peak-to-sidelobe ratio (PSLR). 

In \cite{SA_OFDM}, a subcarrier-aliasing (SA) OFDM  radar system has been studied as an alternative approach to reduce the ADC sampling rate. SA-OFDM radar transmits a signal that only utilizes every \textcolor{black}{$\mu\text{th}$} subcarrier as an active subcarrier, where $\mu$ is determined by the corresponding search algorithm. Then, sub-Nyquist sampling is conducted in the ADC, causing the active subcarriers to alias to the empty subcarrier positions. However, this method leads to a reduction in the maximum unambiguous range that exceeds the reduction factor of the ADC sampling rate.

Finally, a sub-Nyquist sampling (SNS) OFDM radar \cite{SNS_OFDM} system, which transmits a full bandwidth OFDM radar signal, is investigated. The received baseband signals are sub-sampled and folded in the lowest sub-band due to aliasing. The folded signals are restored to the full band signals through an unfolding process with the help of known transmitted symbols. This is done to ensure a high range resolution. However, it also introduces symbol-mismatch noise during the unfolding process. Although it can operate without any additional hardware and does not degrade the range or cause Doppler ambiguities, a complex symbol-mismatch noise cancellation technique that works through iterative noise-canceling processes is required.

In this paper, we introduce phase-coded sub-Nyquist sampling (PC-SNS) OFDM  that reduces the ADC sampling rate without any additional hardware or signal processing. This exploits a phase-coded OFDM waveform with the same structure of SNS-OFDM to cancel the symbol-mismatch noise that occurs during the unfolding process. We describe the proposed waveform structure and its radar processing technique mathematically. Also, we validate the PC-SNS-OFDM radar system through simulations and measurements under various ADC sampling rates.

The remainder of this paper is organized as follows. Section \uppercase\expandafter{\romannumeral2} presents the SNS-OFDM radar system \textcolor{black}{\cite{SNS_OFDM}} and issues that hinder its use. The newly proposed phase-coded OFDM waveform is then introduced to address these issues. Section \uppercase\expandafter{\romannumeral3} demonstrates the effects of phase coding on the radar system with simulation results. Experimental results are presented in Section \uppercase\expandafter{\romannumeral4}, and the paper concludes in Section \uppercase\expandafter{\romannumeral5}.

$Notations$: Boldface variables in lower and upper case symbols represent vectors and matrices, respectively. $\textbf{A} \in \mathbb{C}^{N \times M}$ denotes a complex-valued \textcolor{black}{$N \times M$} matrix $\textbf{A}$. $a_{ij}$ indicates an element of matrix $\textbf{A}$ at the \textcolor{black}{$i\text{th}$} row and the \textcolor{black}{$j\text{th}$} column. $({\cdot})^{T}$ and $({\cdot})^{H}$ denote the transpose and Hermitian transpose operators, respectively. ${\text{diag}({\textbf{a}})}$ denotes a diagonal matrix with diagonal entries of a vector $\textbf{a}$. $\odot$ and $\oslash$ represent the Hadamard (element-wise) product and division operator, respectively.

\section{Signal Model of a Phase-Coded Sub-Nyquist Sampling OFDM Radar System}

In this section, we provide a concise description of the SNS-OFDM radar system\textcolor{black}{\cite{SNS_OFDM}}. Then, the proposed PC-SNS-OFDM waveform is introduced, which aims to overcome the limitations of SNS-OFDM radar. \textcolor{black}{The} signal processing \textcolor{black}{of} this system is then explained in detail. 

\begin{figure}[t!]
    \centering
    \subfloat[]{\includegraphics[scale=0.55]{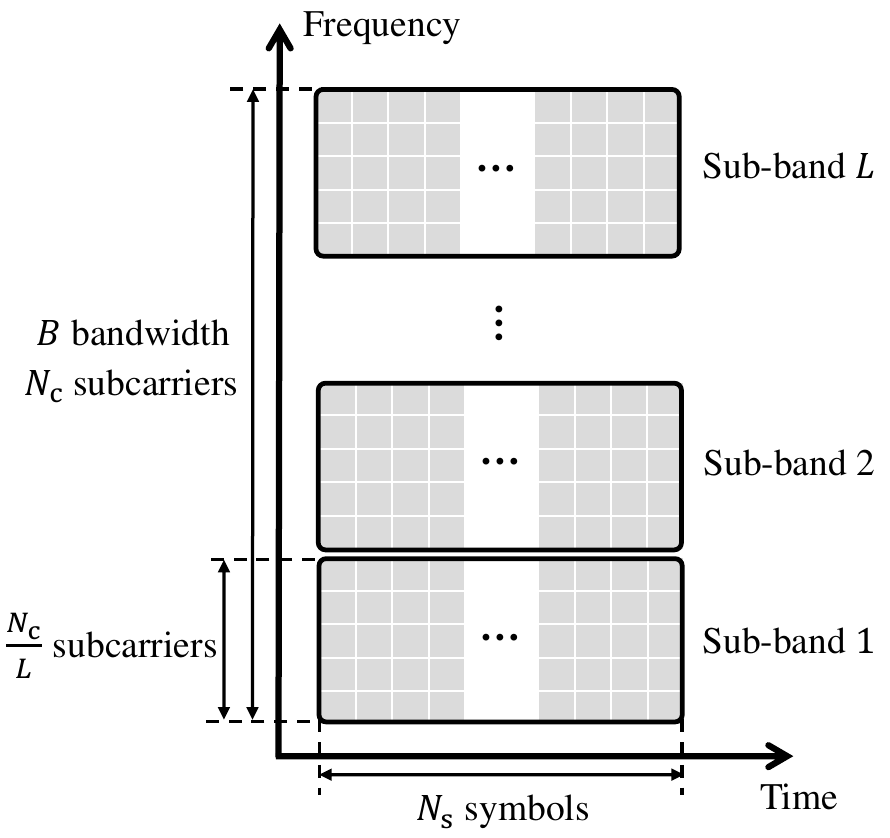}}
    \hfill
    \centering
    \subfloat[]{\includegraphics[scale=0.55]{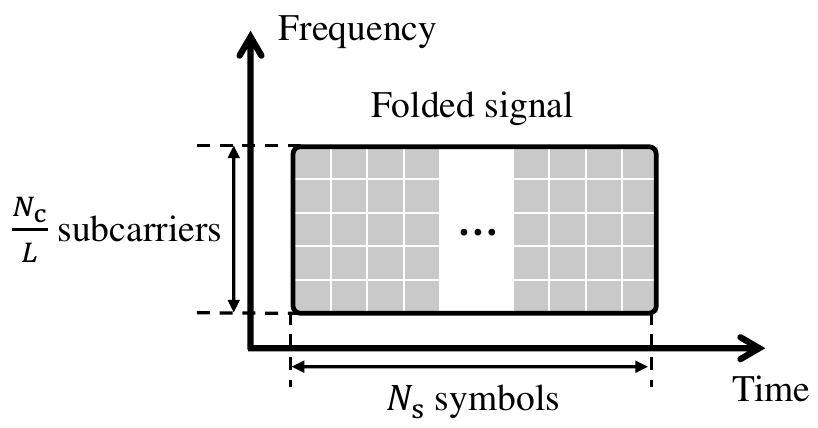}}

    \caption{Frame structure of (a) the received OFDM signal and (b) the folded signal due to sub-Nyquist sampling}
    \label{f1}
\end{figure}

\begin{figure*}[ht!]
    \centering\includegraphics[scale=0.53]{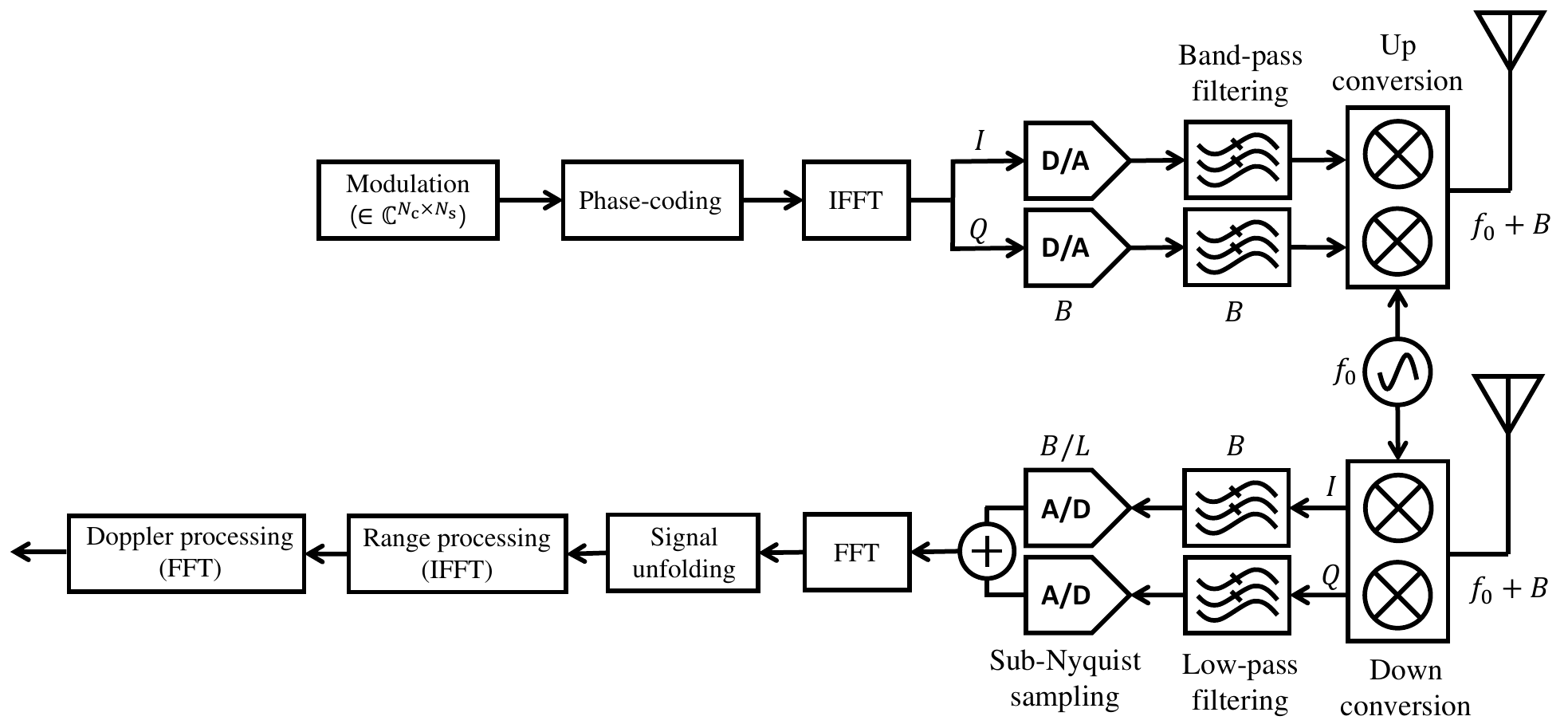}
    \caption{Block diagram of the PC-SNS-OFDM radar system}
    \label{f2}
\end{figure*}

\subsection{Sub-Nyquist sampling OFDM radar}
Given that an OFDM radar signal consists of $N_\text{\textcolor{black}{c}}$ subcarriers and $N_\text{\textcolor{black}{s}}$ OFDM symbols, the transmitted symbol matrix can be expressed as $\textbf{C} \in \mathbb{C}^{N_\text{\textcolor{black}{c}} \times N_\text{\textcolor{black}{s}}}$.
When the transmitted OFDM signal is reflected from a target, the received signal contains information about the range and velocity of the targets. A range steering vector, indicating the ranges from the target to the radar, and a Doppler steering vector, representing the velocities of the target, can be expressed as follows:
\begin{equation}
    \textbf{r}(\tau) = 
    \begin{bmatrix}
        e^{-\text{\textcolor{black}{j}}2 \pi \Delta f \tau} & e^{-\text{\textcolor{black}{j}}2 \pi 2\Delta f \tau} & \cdots & e^{-\text{\textcolor{black}{j}}2 \pi N_\text{\textcolor{black}{c}}\Delta f \tau}
    \end{bmatrix}^{T}\in \mathbb{C}^{N_\text{\textcolor{black}{c}} \times 1},
    \label{e1}
\end{equation}
\begin{equation}
    \textbf{v}(f_\text{\textcolor{black}{D}}) = 
    \begin{bmatrix}
        e^{\text{\textcolor{black}{j}} 2\pi f_\text{\textcolor{black}{D}} T_\text{\textcolor{black}{s}}} & e^{\text{\textcolor{black}{j}} 2\pi 2f_\text{\textcolor{black}{D}} T_\text{\textcolor{black}{s}}} & \cdots & e^{\text{\textcolor{black}{j}} 2\pi N_\text{\textcolor{black}{s}} f_\text{\textcolor{black}{D}} T_\text{\textcolor{black}{s}}}
    \end{bmatrix}^{T} \in \mathbb{C}^{N_\text{\textcolor{black}{s}} \times 1},
    \label{e2}
\end{equation}
where $\tau$ represents the round-trip time to the target and $f_\text{\textcolor{black}{D}}$ denotes the Doppler frequency offset resulting from the relative velocity between the radar and the target. Additionally, $\Delta f$ represents the subcarrier spacing, and $T_\text{\textcolor{black}{s}}$ corresponds to the duration of each OFDM symbol. Assuming that there are $K$ targets, the range matrix and the Doppler matrix can be expanded as 
\begin{equation}
    \textbf{R} = 
    \begin{bmatrix}
        \textbf{r}(\tau_1) & \textbf{r}(\tau_2) & \cdots & \textbf{r}(\tau_K)
    \end{bmatrix}\in \mathbb{C}^{N_\text{\textcolor{black}{c}} \times K},
    \label{e3}
\end{equation}
\begin{equation}
    \textbf{V} = 
    \begin{bmatrix}
        \textbf{v}(f_{\text{\textcolor{black}{D}},1}) & \textbf{v}(f_{\text{\textcolor{black}{D}},2}) & \cdots & \textbf{v}(f_{\text{\textcolor{black}{D}},K})
    \end{bmatrix} \in \mathbb{C}^{N_\text{\textcolor{black}{s}} \times K}.
    \label{e4}
\end{equation}
Also, the attenuation of signals reflected from the targets is expressed as
\begin{equation}
    \textbf{A} = 
    \text{diag}(\alpha_1,\alpha_2, \cdots, \alpha_K) 
    \in \mathbb{C}^{K \times K}.
    \label{e5}
\end{equation}
By using (\ref{e3}), (\ref{e4}), and (\ref{e5}), let $\textbf{X}$ denote the target information matrix as
\begin{equation}
    \textbf{X} = \textbf{R}\textbf{A}\textbf{V}^{T}.
    \label{e6}
\end{equation}

With additive white Gaussian noise $\textbf{W}$, the frequency-domain representation of the received signal can be established as 
\begin{equation}
    \textbf{S} = \textbf{C} \odot \textbf{X} + \textbf{W}, 
    \label{e7}
\end{equation}
\begin{equation}
    \begin{bmatrix}
    \textbf{S}_1 \\ \textbf{S}_2 \\ \vdots \\ \textbf{S}_{L}
    \end{bmatrix}
     = \begin{bmatrix}
    \textbf{C}_1 \\ \textbf{C}_2 \\ \vdots \\ \textbf{C}_{L}
    \end{bmatrix}
    \odot \begin{bmatrix}
    \textbf{X}_1 \\ \textbf{X}_2 \\ \vdots \\ \textbf{X}_{L}
    \end{bmatrix}
    + \begin{bmatrix}
    \textbf{W}_1 \\ \textbf{W}_2 \\ \vdots \\ \textbf{W}_{L}
    \end{bmatrix}.
    \label{e8}
\end{equation}
\textcolor{black}{In this modeling, effects such as range migration and inter-carrier interference (ICI) have been disregarded.} Here, $L$ is a positive proper divisor of the number of subcarriers $N_\text{\textcolor{black}{c}}$, which also corresponds to the number of sub-bands allocated for an OFDM symbol.

Fig. \ref{f1} (a) illustrates the structure of the received OFDM frame. In a conventional OFDM radar system with signal bandwidth $B$, the ADC requires a sampling rate $F_\text{\textcolor{black}{s}}$ greater than $B$ to reconstruct the full-band signal. However, in the SNS-OFDM radar system, the sampling rate $F_\text{\textcolor{black}{s}}$ is set to $B/L$ in the ADC. Consequently, aliasing occurs, resulting in a folded signal, as depicted in Fig. \ref{f1} (b). Let \textbf{Z} be the folded signal of the received symbol of equation (\ref{e8}). 
\begin{equation}
    \textbf{Z} = 
    {\sum_{j=1}^{L}} \, \textbf{S}_j = 
    {\sum_{j=1}^{L}} \,(\textbf{C}_j \odot \textbf{X}_j) + {\sum_{j=1}^{L}} \, \textbf{W}_j
    \label{e9}
\end{equation}

Given that the bandwidth of the folded signal is reduced to $B/L$, the range resolution is also degraded. To overcome this limitation, signal unfolding demodulation is necessary. Let $\textbf{D}_i$ be the result of the element-wise division of $\textbf{Z}$, represented as $\textbf{C}_i$, which is \textcolor{black}{$i\text{th}$} sub-band of unfolded signal $\textbf{D}$. 
\begin{equation}
    \textbf{D} = 
    \begin{bmatrix}
        \textbf{D}_1 \\ \textbf{D}_2 \\ \vdots \\ \textbf{D}_L
    \end{bmatrix}
    = \begin{bmatrix}
        \textbf{Z} \oslash \textbf{C}_1  \\ \textbf{Z} \oslash \textbf{C}_2 \\ \vdots \\ \textbf{Z} \oslash \textbf{C}_L
    \end{bmatrix}
    \label{e10}
\end{equation}
$\textbf{D}$ includes the desired signal containing the range and velocity information of the targets. 

During the signal demodulation process, additional symbol-mismatch terms are introduced due to other sub-band components in \textcolor{black}{\textbf{D}}. The presence of symbol-mismatch noise and folded white noise significantly reduces the SNR of the targets in the SNS-OFDM radar system. As a result, symbol-mismatch cancellation processing is proposed in order to mitigate the symbol-mismatch noise in \cite{SNS_OFDM}. In the following subsection, a \textcolor{black}{PC-SNS-OFDM} radar system is proposed to reduce the symbol-mismatch noise without the need for complex signal recovery processing.

\begin{figure}[t!]
    \centering\includegraphics[scale=0.55, angle = 270]{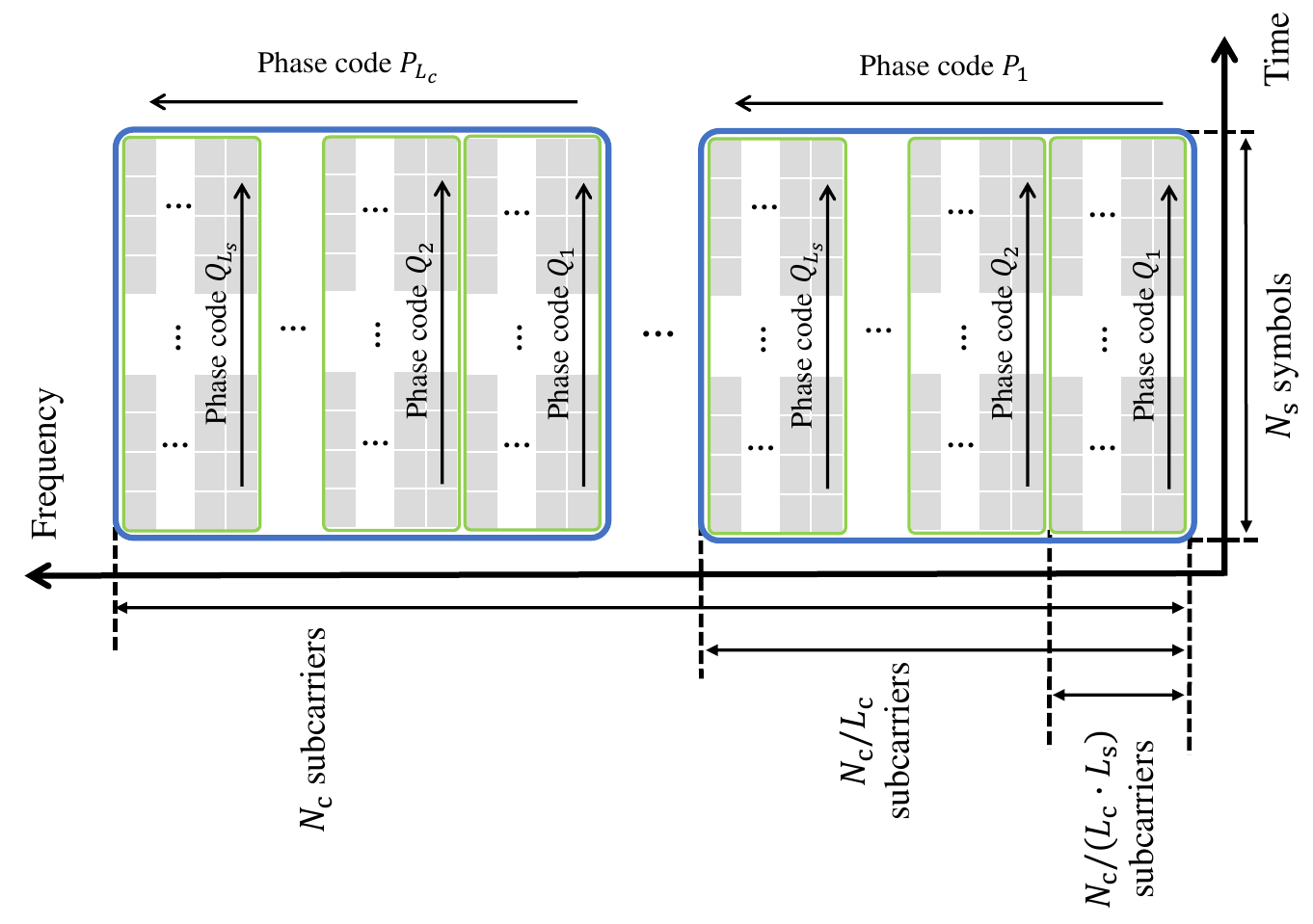}
    \caption{Phase-coded OFDM waveform structure with $L_\text{\textcolor{black}{c}}$ sub-bands encoded in the frequency domain and $L_\text{\textcolor{black}{s}}$ sub-bands encoded in the time domain}
    \label{f3}
\end{figure}

\subsection{Phase-coded sub-Nyquist sampling OFDM radar}
Fig. \ref{f2} presents a block diagram of the proposed PC-SNS-OFDM radar system. Similarly to the SNS-OFDM radar system \textcolor{black}{\cite{SNS_OFDM}}, the basic structure of the system involves receiving a folded signal from the ADC through uniform sub-Nyquist sampling and then unfolding the signal to restore it to the original full band signal with subsequent radar processing. However, unlike the SNS-OFDM radar system that employs conventional OFDM signals, the PC-SNS-OFDM radar system utilizes phase-coded OFDM waveforms in both the time and frequency domains.

First, the transmitted symbol is divided into $L$ sub-bands, where all $L$ sub-bands employ the same modulation symbol. In this section, the modulation symbol of the sub-bands is defined as $\textbf{C}_\text{\textcolor{black}{sub}}$
\begin{equation}
\textbf{C}_\text{\textcolor{black}{sub}} = 
\begin{bmatrix}
    c_{11} & c_{21} & \cdots & c_{N_{s}1} \\
    c_{12} & c_{22} & \cdots & c_{N_{s}2} \\
    \vdots & \vdots & \ddots & \vdots \\
    c_{1 (N_\text{\textcolor{black}{c}}/{L})} & c_{2 ({N_\text{\textcolor{black}{c}}}/{L})} & \cdots & c_{N_\text{\textcolor{black}{s}}({N_\text{\textcolor{black}{c}}}/{L})}
\end{bmatrix}
\in \mathbb{C}^{\frac{N_\text{\textcolor{black}{c}}}{L} \times N_\text{\textcolor{black}{s}}}
\label{e11}
\end{equation}
Similarly, $L$ is the positive proper divisor of the number of subcarriers $N_\text{\textcolor{black}{c}}$. For the orthogonality of these sub-bands, a total of $L$ phase codes are required. Let the number of the phase codes set in the time domain be denoted as $L_\text{\textcolor{black}{s}}$ and that in the frequency domain be denoted as $L_\text{\textcolor{black}{c}}$. Then, the relationship between $L$ can be established as  
\begin{equation}
    L = \textcolor{black}{L_\text{\textcolor{black}{c}} \cdot L_\text{\textcolor{black}{s}}}.  
    \label{e12}
\end{equation}

Fig. \ref{f3} illustrates the waveform structure of the proposed phase-coded OFDM waveform. The waveform consists of two levels of sub-band division: $L_\text{\textcolor{black}{c}}$ sub-bands in the first level and $L_\text{\textcolor{black}{s}}$ sub-bands in the second level. $L_\text{\textcolor{black}{c}}$ sub-bands in the first level are established using orthogonal phase codes in the frequency domain. Subsequently, each of these sub-bands is further divided into $L_\text{\textcolor{black}{s}}$ sub-bands in the second level, providing orthogonality through the application of phase codes in the time domain. 

To represent the proposed waveform mathematically, the \textcolor{black}{$q\text{th}$} phase code in the time domain can be defined as follows:
\begin{equation}
    \textbf{Q}_q
    = \begin{bmatrix}
        e^{\text{\textcolor{black}{j}}2 \pi q /L_\text{\textcolor{black}{s}}} & e^{\text{\textcolor{black}{j}}2 \pi q2 /L_\text{\textcolor{black}{s}}} &  \cdots & e^{\text{\textcolor{black}{j}}2 \pi q N_\text{\textcolor{black}{s}} /L_\text{\textcolor{black}{s}}} \\
        \vdots & \vdots & \ddots & \vdots \\
        e^{\text{\textcolor{black}{j}}2 \pi q /L_\text{\textcolor{black}{s}}} & e^{\text{\textcolor{black}{j}}2 \pi q2 /L_\text{\textcolor{black}{s}}} &  \cdots & e^{\text{\textcolor{black}{j}}2 \pi q N_\text{\textcolor{black}{s}} /L_\text{\textcolor{black}{s}}}
    \end{bmatrix}\in \mathbb{C}^{\frac{N_\text{\textcolor{black}{c}}}{L_\text{\textcolor{black}{c}} L_\text{\textcolor{black}{s}}} \times N_\text{\textcolor{black}{s}}}.
    \label{e13}
\end{equation}

To define the \textcolor{black}{$p\text{th}$} frequency-domain phase code, first, we divide the \textcolor{black}{$p\text{th}$} code matrix into $L_\text{\textcolor{black}{s}}$ submatrices along the frequency axis. The first submatrix $\textbf{P}_p^{(1)}$ is initially defined as 
\begin{equation}
    \textbf{P}_p^{(1)}
    = \begin{bmatrix}
        e^{\text{\textcolor{black}{j}}2 \pi p /L_\text{\textcolor{black}{c}}} &  \cdots & e^{\text{\textcolor{black}{j}}2 \pi p /L_\text{\textcolor{black}{c}}} \\
        e^{\text{\textcolor{black}{j}}2 \pi p 2/L_\text{\textcolor{black}{c}}}  & \cdots & e^{\text{\textcolor{black}{j}}2 \pi p 2/L_\text{\textcolor{black}{c}}} \\
        \vdots & \ddots & \vdots \\
        e^{\text{\textcolor{black}{j}}2 \pi p (N_\text{\textcolor{black}{c}}/L) /L_\text{\textcolor{black}{c}}}  & \cdots & e^{\text{\textcolor{black}{j}}2 \pi p (N_\text{\textcolor{black}{c}}/L) /L_\text{\textcolor{black}{c}}} 
    \end{bmatrix} \in \mathbb{C}^{\frac{N_\text{\textcolor{black}{c}}}{L_\text{\textcolor{black}{c}} L_\text{\textcolor{black}{s}}} \times N_\text{\textcolor{black}{s}}}.
    \label{e14}
\end{equation}
At this stage, we introduce the variable $\phi$ to define the other submatrices in the \textcolor{black}{$p\text{th}$} code matrix. This is determined as
\begin{equation}
    \phi_p = e^{\text{\textcolor{black}{j}}2 \pi p (N_\text{\textcolor{black}{c}}/L) /L_\text{\textcolor{black}{c}}}.
    \label{e15}
\end{equation}
Then, the following relationship is established between $\textbf{P}_p^{(1)}$ and $\textbf{P}_p^{(q)}$ as
\begin{equation}
    \textbf{P}_p^{(q)} = \phi_p^{q-1} \cdot \textbf{P}_p^{(1)}.
    \label{e16}
\end{equation}
The \textcolor{black}{$p\text{th}$} frequency-domain phase code can be defined as a block matrix of $L_\text{\textcolor{black}{s}}$ submatrices,
\begin{equation}
    \textbf{P}_p = \begin{bmatrix}
        \textbf{P}_p^{(1)} \\ \textbf{P}_p^{(2)} \\ \vdots \\ \textbf{P}_p^{(L\text{\textcolor{black}{s}})}
    \end{bmatrix} \in \mathbb{C}^{\frac{N_\text{\textcolor{black}{c}}}{L_\text{\textcolor{black}{c}}} \times N_\text{\textcolor{black}{s}}}.
    \label{e17}
\end{equation}
Using the time-frequency phase code defined above, the phase-coded waveform $\textbf{C}_{pq}$ of the \textcolor{black}{$p\text{th}$} sub-band at the first level and the \textcolor{black}{$q\text{th}$} sub-band at the second level can be expressed as follows:
\begin{equation}
    \textbf{C}_{pq}
    = \textbf{C}_\text{\textcolor{black}{sub}} \odot \textbf{P}_p^{(q)} \odot \textbf{Q}_q \in \mathbb{C}^{\frac{N_\text{\textcolor{black}{c}}}{L_\text{\textcolor{black}{c}} L_\text{\textcolor{black}{s}}} \times N_\text{\textcolor{black}{s}}}.
    \label{e18}
\end{equation}
The entire transmitted phase-coded OFDM signal \textbf{C} can be represented as a block matrix augmented by \textcolor{black}{$L_\text{\textcolor{black}{c}} \cdot L_\text{\textcolor{black}{s}}$} submatrices, expressed as shown below.
\begin{equation}
    \textbf{C}
    = \begin{bmatrix}
    \textbf{C}_{11} \\ \textbf{C}_{12} \\ \vdots \\  \textbf{C}_{L_\text{\textcolor{black}{c}} L_\text{\textcolor{black}{s}}} 
    \end{bmatrix}\in \mathbb{C}^{N_\text{\textcolor{black}{c}} \times N_\text{\textcolor{black}{s}}}.
    \label{e19}
\end{equation}
The received signal matrix is represented as 
\begin{equation}
    \textbf{S} = \textbf{C} \odot \textbf{X} + \textbf{W}.
    \label{e20}
\end{equation}
\textcolor{black}{Here, $\textbf{X}_{pq}$ and $\textbf{W}_{pq}$ are the target information and white Gaussian noise corresponding to the sub-band phase-coded waveform $\textbf{C}_{pq}$, respectively.}
Equation (\ref{e20}) can be expressed in an alternative form as
\begin{equation}
    \begin{bmatrix}
        \textbf{S}_{11} \\ \textbf{S}_{12} \\ \vdots \\  \textbf{S}_{L_\text{\textcolor{black}{c}}L_\text{\textcolor{black}{s}}}
    \end{bmatrix} = 
    \begin{bmatrix}
        \textbf{C}_{11} \\ \textbf{C}_{12} \\ \vdots \\  \textbf{C}_{L_\text{\textcolor{black}{c}}L_\text{\textcolor{black}{s}}}
    \end{bmatrix} \odot
    \begin{bmatrix}
        \textbf{X}_{11} \\ \textbf{X}_{12} \\ \vdots \\  \textbf{X}_{L_\text{\textcolor{black}{c}}L_\text{\textcolor{black}{s}}}
    \end{bmatrix} +
    \begin{bmatrix}
        \textbf{W}_{11} \\ \textbf{W}_{12} \\ \vdots \\  \textbf{W}_{L_\text{\textcolor{black}{c}} L_\text{\textcolor{black}{s}}}
    \end{bmatrix}.
    \label{e21}
\end{equation}

In the PC-SNS-OFDM radar system, the sampling rate of the ADC is also set to $B/L$, where $B$ represents the bandwidth and $L$ denotes the number of sub-bands. This sampling rate leads to aliasing, causing the formation of folded signals in the $L$ sub-bands. In the PC-SNS-OFDM scheme, the folded signal can be expressed in the frequency domain as follows.
\begin{equation}
    \textbf{Z} = {\sum_{p=1}^{L_\text{\textcolor{black}{c}}}}{\sum_{q=1}^{L_\text{\textcolor{black}{s}}}} \,(\textbf{C}_{pq} \odot \textbf{X}_{pq}) + {\sum_{p=1}^{L_\text{\textcolor{black}{c}}}}{\sum_{q=1}^{L_\text{\textcolor{black}{s}}}} \, \textbf{W}_{pq}
    \label{e22}
\end{equation}
Let $\textbf{D}_{pq}$ represent a matrix obtained from the element-wise division of folded signal $\textbf{C}_{pq}$; the unfolded signal can be expressed as
\begin{gather}
    \textbf{D} = 
    \begin{bmatrix}
        \textbf{D}_{11} \\ \textbf{D}_{12} \\ \vdots  \\ \textbf{D}_{L_\text{\textcolor{black}{c}} L_\text{\textcolor{black}{s}}}
    \end{bmatrix} =
    \begin{bmatrix}
        \textbf{Z} \oslash \textbf{C}_{11} \\ \textbf{Z} \oslash \textbf{C}_{12} \\ \vdots  \\ \textbf{Z} \oslash \textbf{C}_{L_\text{\textcolor{black}{c}} L_\text{\textcolor{black}{s}}}
    \end{bmatrix} 
    \\ =
    \begin{bmatrix}
        \textbf{X}_{11} \\ \textbf{X}_{12} \\ \vdots \\ \textbf{X}_{L_\text{\textcolor{black}{c}} L_\text{\textcolor{black}{s}}}
    \end{bmatrix} \odot
    \begin{bmatrix}
        {\sum_{p, q}} \, (\textbf{C}_{pq} \oslash \textbf{C}_{11}) \\ {\sum_{p, q}} \, (\textbf{C}_{pq} \oslash \textbf{C}_{12}) \\ \vdots \\ {\sum_{p, q}} \, (\textbf{C}_{pq} \oslash \textbf{C}_{L_\text{\textcolor{black}{c}} L_\text{\textcolor{black}{s}}})
    \end{bmatrix} + 
    \begin{bmatrix}
        {\sum_{p, q}} \, \textbf{W}_{pq} \oslash \textbf{C}_{11} \\
        {\sum_{p, q}} \, \textbf{W}_{pq} \oslash \textbf{C}_{12} \\ \vdots
        \\ {\sum_{p, q}} \, \textbf{W}_{pq} \oslash \textbf{C}_{L_\text{\textcolor{black}{c}} L_\text{\textcolor{black}{s}}}
    \end{bmatrix}.
    \label{e23}
\end{gather}
Equation (\ref{e23}) can be represented simply as
\begin{equation}
    \textbf{D} = \textbf{X} \odot \textbf{Y} + \textbf{W}_\text{F}.
    \label{e24}
\end{equation}
The crucial aspect to note here is \textbf{Y}. In the conventional SNS-OFDM system, \textbf{Y} becomes a matrix with random symbol entries, leading to symbol-mismatch noise. However, in the PC-SNS-OFDM system, ambiguous peaks arise instead due to the periodicity of the phase-coded waveforms. To address the effect of \textbf{Y}, it is necessary to denote the relationship of time-frequency phase codes in (\ref{e13}), (\ref{e14}), and (\ref{e16}).
\begin{equation}
    \textbf{P}_p = \textbf{P}_{p-1} \odot \textbf{P}_1
    \label{e25}
\end{equation}
\begin{equation}
    \textbf{Q}_q = \textbf{Q}_{q-1} \odot \textbf{Q}_1
    \label{e26}
\end{equation}
\begin{equation}
    \textbf{P}_{L_\text{\textcolor{black}{c}}+p} = \textbf{P}_{p}
    \label{e27}
\end{equation}
\begin{equation}
    \textbf{Q}_{L_\text{\textcolor{black}{s}}+q} = \textbf{Q}_{q}
    \label{e28}
\end{equation}
Using the above relationships, the \textcolor{black}{first} block of \textbf{Y} can be simply reduced as follows. 
\textcolor{black}{\begin{align}
    {\sum_{p,q}} \, (\textbf{C}_{pq} \oslash \textbf{C}_{11}) 
    & = {\sum_{p,q}} \, (\textbf{P}_{p}^{(q)} \oslash \textbf{P}_{1}^{(1)} ) \odot (\textbf{Q}_{q} \oslash \textbf{Q}_{1} ) \\
    & = {\sum_{p,q}} \, \textbf{P}_{p-1}^{(q)} \odot \textbf{Q}_{q-1} \\
    & = {\sum_{p,q}} \, \textbf{P}_{p}^{(q)} \odot \textbf{Q}_{q}
    \label{e29}
\end{align}
(\ref{e29}) is established because of $\textbf{P}_{0}^{(q)} = \textbf{P}_{L_\text{\textcolor{black}{c}}}^{(q)} $ and $\textbf{Q}_{0} = \textbf{Q}_{L_\text{\textcolor{black}{s}}}$.}
Similarly, the other submatrix block of \textbf{Y} can also be calculated as 
\textcolor{black}{\begin{align}
    {\sum_{p,q}} \, (\textbf{C}_{pq} \oslash \textbf{C}_{ab}) 
    & = {\sum_{p,q}} \, \textbf{P}_{p-a}^{(q)} \odot \textbf{Q}_{q-b} \\
    & = {\sum_{p,q}} \, \textbf{P}_{p}^{(q)} \odot \textbf{Q}_{q}
    \label{e29-1}
\end{align}
where $a \in 1, 2, \cdots, L_\text{c}$ and  $b \in 1, 2, \cdots, L_\text{s}$. Therefore, \textbf{Y} can be expressed as}
\begin{equation}
    \begin{bmatrix}
        {\sum_{p,q}} \, (\textbf{C}_{pq} \oslash \textbf{C}_{11}) \\ {\sum_{p,q}} \, (\textbf{C}_{pq} \oslash \textbf{C}_{12}) \\ \vdots \\ {\sum_{p,q}} \, (\textbf{C}_{pq} \oslash \textbf{C}_{L_\text{\textcolor{black}{c}} L_\text{\textcolor{black}{s}}})
        \end{bmatrix} = \begin{bmatrix}
        {\sum_{p,q}}\textbf{P}_{p}^{(q)} \odot \textbf{Q}_{q} \\
        {\sum_{p,q}}\textbf{P}_{p}^{(q)} \odot \textbf{Q}_{q} \\
        \vdots \\ {\sum_{p,q}}\textbf{P}_{p}^{(q)} \odot \textbf{Q}_{q}
    \end{bmatrix}.
    \label{e30}
\end{equation}

In the OFDM radar system, the target's range manifests as a phase shift on the frequency axis, while the target's velocity manifests as a phase shift on the time axis. Consequently, $\textbf{P}_{p}^{(q)}$ represents a phase code in the frequency domain, providing a virtual time delay effect. Then, $\textbf{P}_{p}^{(q)}$ has a virtual range of $\Delta R$, and the following equation holds:
\begin{equation}
    e^{-\text{\textcolor{black}{j}}2 \pi \Delta f (2 \Delta R/{c})} = e^{\text{\textcolor{black}{j}}2 \pi p / L_\text{\textcolor{black}{c}}}.
    \label{e31}
\end{equation}
It is evident that the phase difference between two adjacent elements along the frequency axis in matrix $\textbf{P}_{p}^{(q)}$ periodically duplicates the desired target peak along the range axis. Therefore, the ambiguous range peaks are located at
\begin{equation}
    r_\text{\textcolor{black}{amb}} = \left( r_k + \frac{r_\text{\textcolor{black}{max}}}{L_\text{\textcolor{black}{c}}}p \right) \text{mod} \,\, r_\text{\textcolor{black}{max}}, \quad  p = 1, 2, \cdots, L_\text{\textcolor{black}{c}}-1,
    \label{e32}
\end{equation}
where $r_k$ is the actual range of the $k\text{th}$ target and $r_\text{\textcolor{black}{max}}$ is the maximum detectable range of the conventional OFDM radar system.
Similarly, $\textbf{Q}_{q}$ represents the phase code in the time domain, resulting in a virtual relative velocity effect. If we set $\Delta V$ as the virtual relative velocity of $\textbf{Q}_{q}$, the following formula is established:
\begin{equation}
    e^{\text{\textcolor{black}{j}}2 \pi (2 \Delta V/{\lambda}) T_\text{\textcolor{black}{s}}} = e^{\text{\textcolor{black}{j}}2 \pi q / L_\text{\textcolor{black}{s}}}.
    \label{e33}
\end{equation}
Again, the phase difference between two adjacent elements along the time axis in matrix $\textbf{Q}_{q}$ produces periodic ambiguous peaks along the Doppler axis. Using (\ref{e33}), the ambiguous velocity peak positions are determined as 
\begin{equation}
    v_\text{\textcolor{black}{amb}}= \left(v_k + \frac{v_\text{\textcolor{black}{max}}}{L_\text{\textcolor{black}{s}}}q\right) \text{mod} \,\, v_\text{\textcolor{black}{max}},  \quad q = 1, 2, \cdots, L_\text{\textcolor{black}{s}}-1,
    \label{e34}
\end{equation}
where $v_k$ is the actual relative velocity of \textcolor{black}{$k\text{th}$} target, $\lambda$ is the wavelength of the radar signal and $v_\text{\textcolor{black}{max}}$ represents the maximum detectable velocity of the conventional OFDM radar system.

\begin{figure}[t!]
    \centering
    \subfloat[]{\includegraphics[scale=0.47]{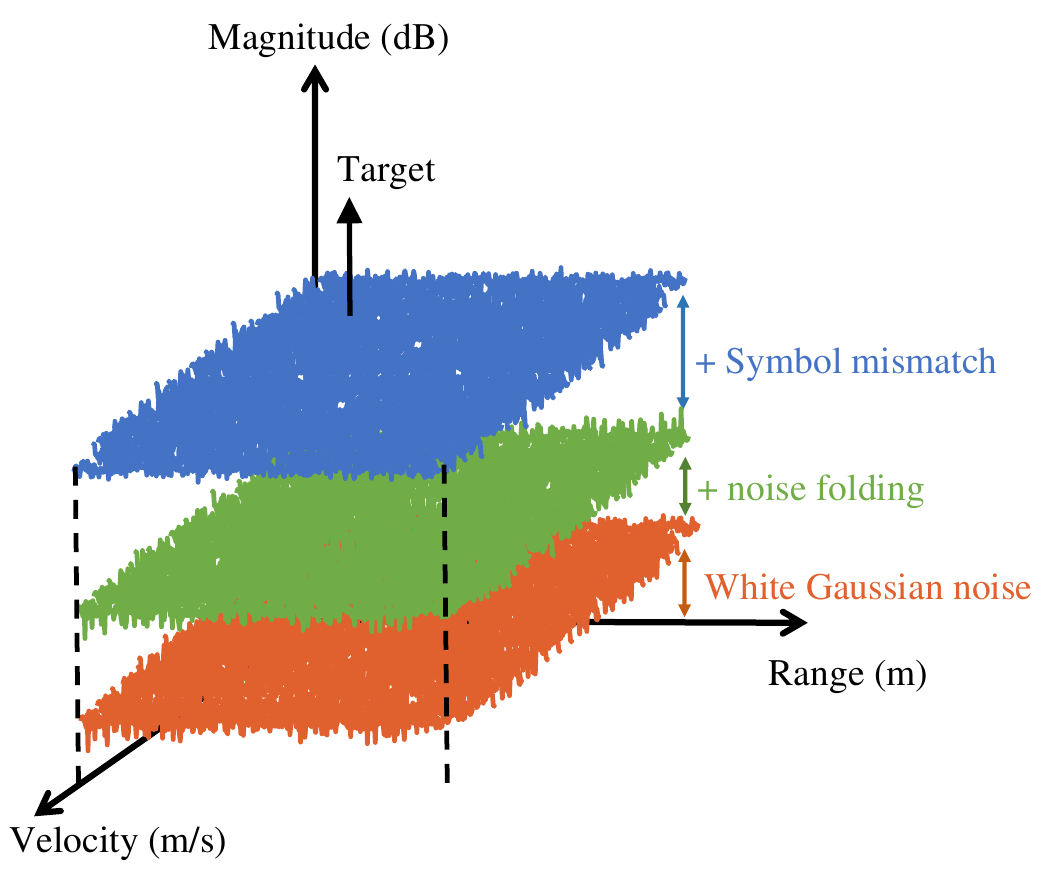}}
    
    \centering
    \subfloat[]{\includegraphics[scale=0.47]{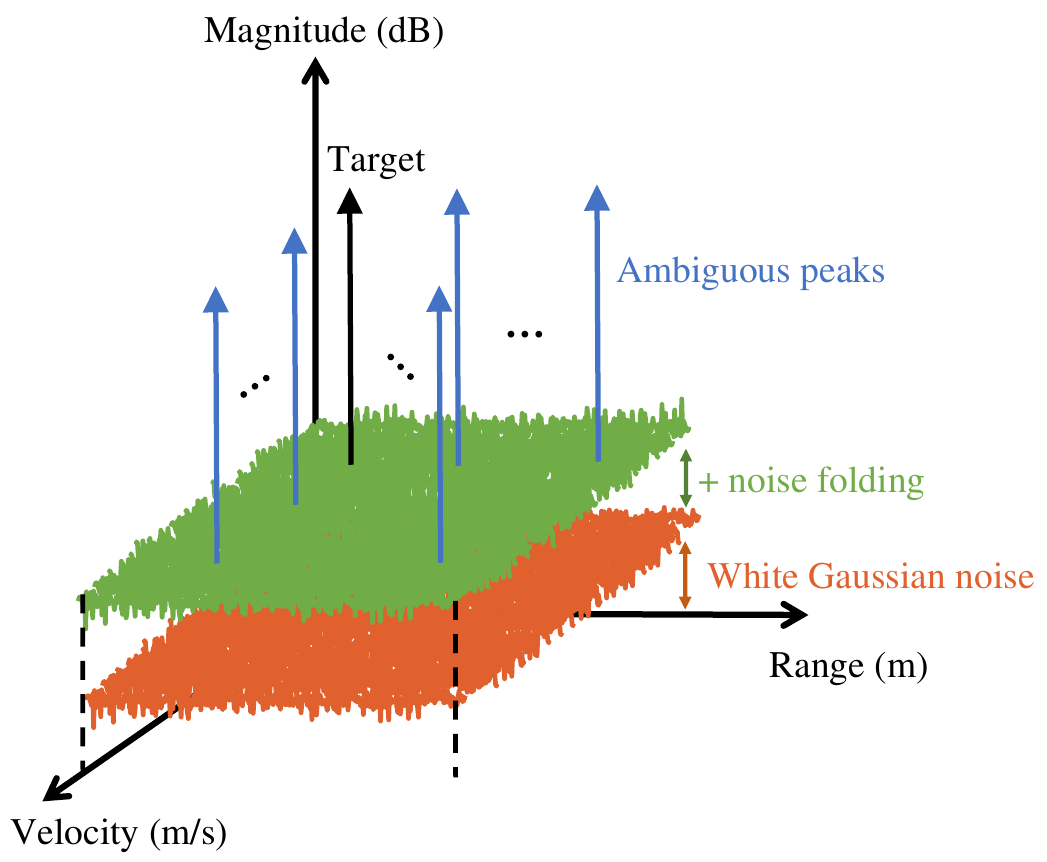}}

    \caption{Noise levels of (a) the SNS-OFDM radar system, and (b) the proposed PC-SNS-OFDM radar system}
    \label{f6}
\end{figure}

Hence, $\textbf{X} \odot \textbf{Y}$ generates a total of $L$ peaks per target, forming $L_\text{\textcolor{black}{c}}$ peaks at intervals of $r_\text{\textcolor{black}{max}}/{L_\text{\textcolor{black}{c}}}$ along the range axis and forming $L_\text{\textcolor{black}{s}}$ peaks at intervals of $v_\text{\textcolor{black}{max}}/{L_\text{\textcolor{black}{s}}}$ along the velocity axis. Among these peaks, one represents the actual target's range and velocity, while the $L-1$ peaks correspond to ambiguous peaks. In this system, the maximum unambiguous range and velocity can be denoted as 
\begin{equation}
    r_\text{\textcolor{black}{u}} = \frac{r_\text{\textcolor{black}{max}}}{L_\text{\textcolor{black}{c}}},
    \label{e35}
\end{equation}
and
\begin{equation}
    v_\text{\textcolor{black}{u}} = \frac{v_\text{\textcolor{black}{max}}}{L_\text{\textcolor{black}{s}}}.
    \label{e36}
\end{equation}
Fig. \ref{f6} illustrates a conceptual diagram depicting the comparison of the noise levels between the SNS-OFDM\textcolor{black}{\cite{SNS_OFDM}} and PC-SNS-OFDM systems. First, it can be observed that SNS-OFDM radar introduces noise folding and symbol-mismatch noise, resulting in a substantial increase in the noise level within the noise floor of the conventional OFDM radar system. However, in the PC-SNS-OFDM radar system, while there is an increase in the noise level due to noise folding, symbol-mismatch noise is absent. Instead, range-Doppler ambiguous peaks arise, which in turn reduce the maximum unambiguous range and velocity of the radar.

\begin{figure*}[t!]
    \centering
    \subfloat[]{\includegraphics[scale=0.545]{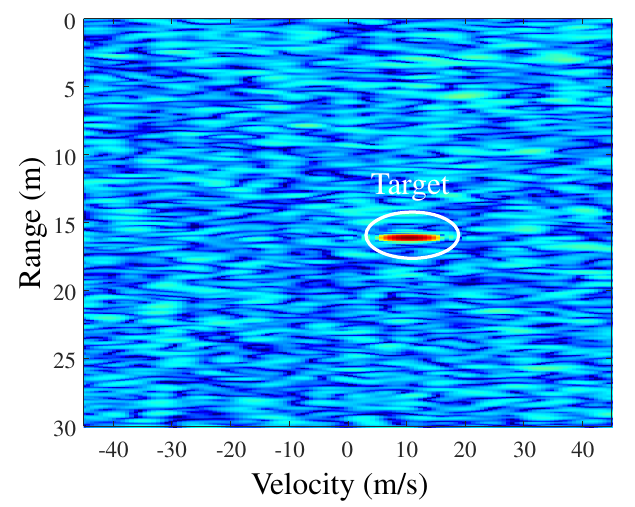}} 
    \hfill
    \centering
    \subfloat[]{\includegraphics[scale=0.545]{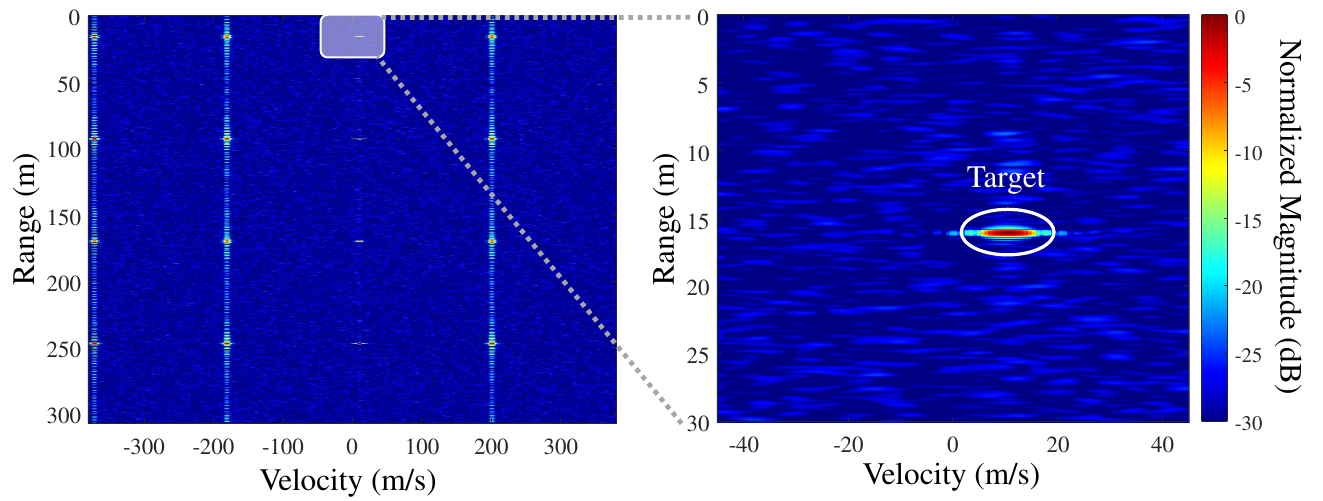}}
    \caption{Range-Doppler maps of (a) the SNS-OFDM radar system ($L = 16$) and (b) the proposed PC-SNS-OFDM radar system (\textcolor{black}{$L_\text{\textcolor{black}{c}} = 4$,  $L_\text{\textcolor{black}{s}} = 4$,  $L = 16$})}
    \label{f7}
\end{figure*}

\begin{figure}[t!]
    \centering
    \subfloat[]{\includegraphics[scale=0.55]{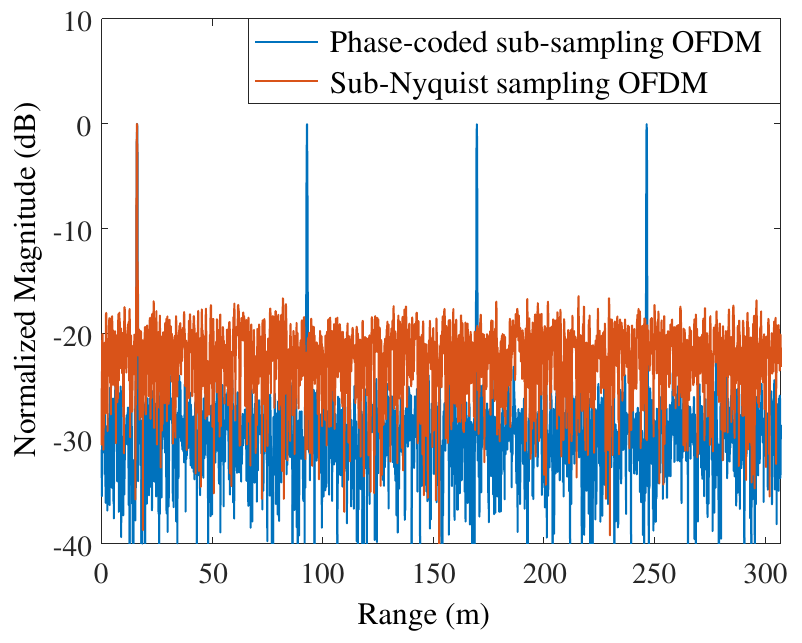}}
    
    \centering
    \subfloat[]{\includegraphics[scale=0.55]{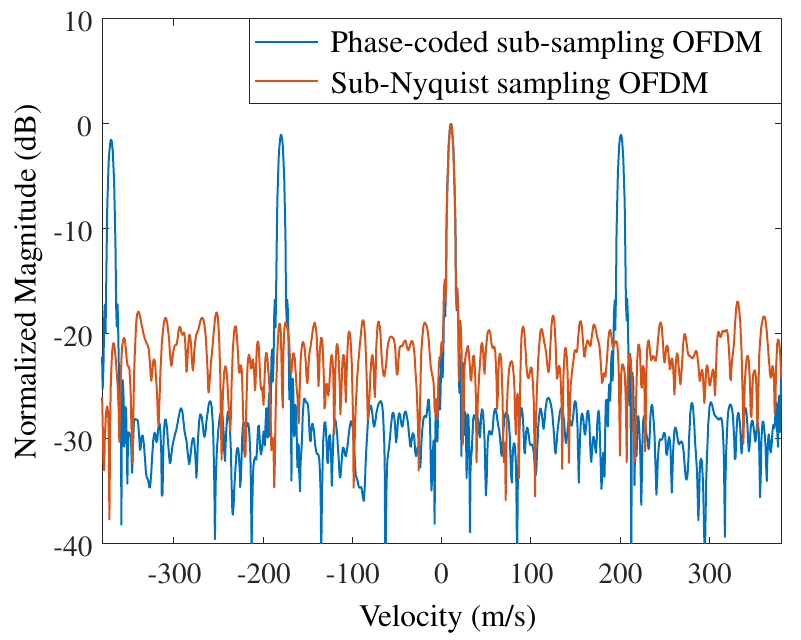}}

    \caption{(a) Range and (b) Doppler profiles of the SNS-OFDM radar system ($L = 16$) and the proposed PC-SNS-OFDM radar system ($L_\text{\textcolor{black}{c}} = 4$,  $L_\text{\textcolor{black}{s}} = 4$,  $L = 16$)}
    \label{f8}
\end{figure}

\begin{figure}[t!]
    \centering
    \subfloat[]{\includegraphics[scale=0.55]{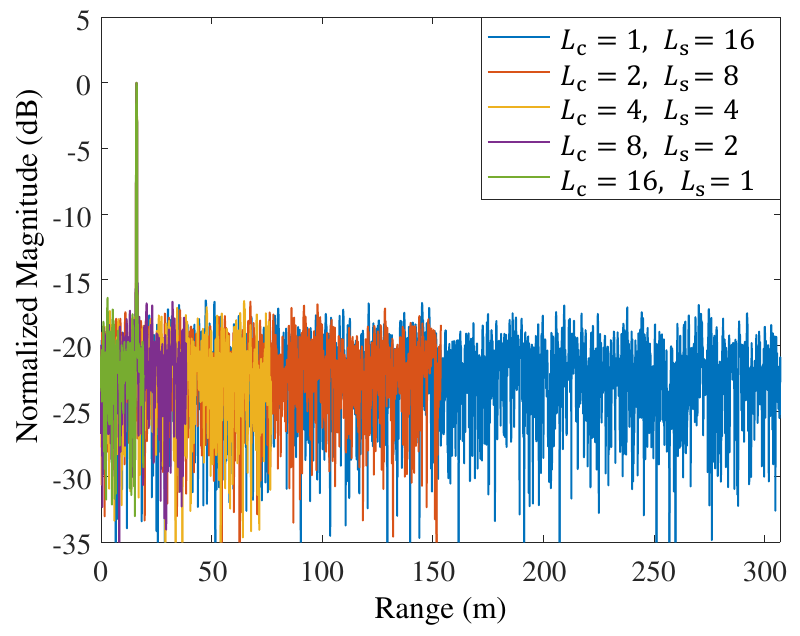}}
    
    \centering
    \subfloat[]{\includegraphics[scale=0.55]{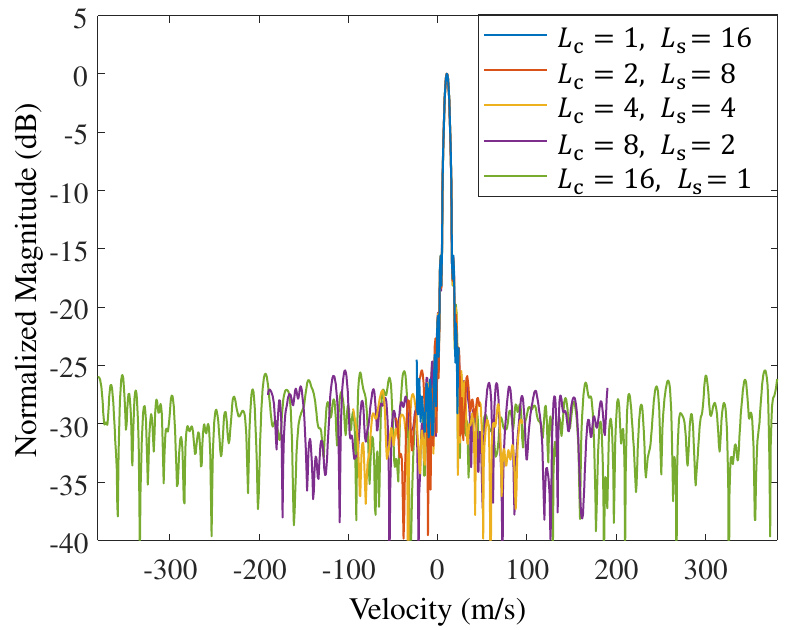}}

    \caption{(a) Range and (b) Doppler profiles of the proposed PC-SNS-OFDM radar system at $L=16$}
    \label{f9}
\end{figure}

\section{Analysis of PC-SNS-OFDM Radar System}
This section presents simulation results of the proposed PC-SNS-OFDM radar system under various conditions. Also, we compare the proposed radar system with state-of-the-art OFDM radar systems that aim to reduce ADC sampling rates.

\begin{table}[t!]
\centering
\fontsize{8.3}{14}\selectfont
\caption{OFDM radar parameters for the simulations}
\label{t1}
\begin{tabular}{>{\centering\arraybackslash} m{20em} | >{\centering\arraybackslash} m{7em}}
    \toprule
    Parameters    & Values\\ \hline
    \midrule
Number of subcarriers, $N_\text{\textcolor{black}{c}}$           & 2048         \\ \hline
Number of symbols, $N_\text{\textcolor{black}{s}}$               & 256         \\ \hline
Subcarrier spacing, $\Delta f$                & 488 kHz     \\ \hline
Total signal bandwidth, B                       & 1 GHz     \\ \hline
OFDM symbol duration, $T_\text{\textcolor{black}{sym}}$    & 2.05 \textcolor{black}{$\upmu$}s    \\ \hline
CP duration,  $T_\text{\textcolor{black}{CP}}$                   & 0.51 \textcolor{black}{$\upmu$}s \\ \hline
OFDM symbol duration with CP, $T_\text{\textcolor{black}{s}}$   & 2.56  \textcolor{black}{$\upmu$}s \\ \hline
Carrier frequency, $f_\text{0}$               & 77 GHz  \\
    \bottomrule
\end{tabular}
\end{table}

\subsection{Simulation results}

Table \ref{t1} summarizes the parameters of the OFDM signal utilized in the simulation. The signal consists of 2048 subcarriers and 256 OFDM symbols. The bandwidth of the signal is 1 GHz, and the subcarrier spacing is set to 488 kHz. The carrier frequency is fixed at 77 GHz. From the aforementioned parameters, the range resolution is calculated to be 15 cm, while the velocity resolution is determined to be 2.96 m/s. Additionally, the simulation establishes a maximum detectable range of 307.2 m and a maximum detectable velocity of $\pm$380.5 m/s. For the purpose of elucidating the range and Doppler ambiguity effects, the simulation assumes a single target positioned 16 m from the radar with a velocity of 10 m/s. The noise level for this simulation is set to \textcolor{black}{$-$20} dBm.

Fig. \ref{f7} presents the simulated results of range-Doppler maps comparing the SNS-OFDM\textcolor{black}{\cite{SNS_OFDM}} and the PC-SNS-OFDM radar systems. The simulation is conducted with $L$ = 16 and an ADC sampling rate of $B/L = 62.5$ MHz. A phase-coded waveform is generated with $L_\text{\textcolor{black}{c}}$ = 4 and $L_\text{\textcolor{black}{s}}$ = 4. In Fig. 5 (a), the graph represents the range-Doppler map of SNS-OFDM, while Fig. 5(b) displays the range-Doppler maps of PC-SNS-OFDM. The noise levels of Fig. 5 (a) and (b) are observed to be \textcolor{black}{$-$22.1} dB and \textcolor{black}{$-$30.5} dB, respectively, with the PC-SNS-OFDM radar system exhibiting a lower noise level by approximately 8 dB. The range-Doppler maps provide an overview of the results across the entire range bin and velocity bin. As depicted in the first graph of (b), 16 peaks occur on both the range axis and velocity axis. Consequently, these ambiguous peaks lead to reductions in the maximum unambiguous range and velocity: $r_\text{\textcolor{black}{u}}$ to 76.8 m and $v_\text{\textcolor{black}{u}}$ to $\pm$95.125 m/s, respectively.

\begin{table*}[t!]
\centering
\fontsize{9}{11.5}\selectfont
\caption{Comparisons of the state-of-the-art OFDM radar systems with reduced baseband sampling rates}
\label{t2}
\renewcommand{\arraystretch}{1.5}
\begin{tabular}{>{\centering\arraybackslash} m{9em} | >{\centering\arraybackslash} m{8.5em}| >{\centering\arraybackslash} m{8.5em}| >{\centering\arraybackslash} m{8em}| >{\centering\arraybackslash} m{8em}| >{\centering\arraybackslash} m{7em}}
    \toprule
    & SC-OFDM \cite{SC_OFDM2} & FC-OFDM \cite{FC_OFDM} & SA-OFDM$^{*}$ \cite{SA_OFDM} & SNS-OFDM \cite{SNS_OFDM}& \makecell{PC-SNS-OFDM \\ {(This work)}} \\ 
    \hline
    \midrule
ADC sampling rate  & ${B}/{L}$ & ${B}/{L}$ & ${B}/{L}$ & ${B}/{L}$ & \makecell{${B}/{L}$ \\ (${L} = {L_\text{\textcolor{black}{c}}}\cdot{L_\text{\textcolor{black}{s}}}$)}   \\ \hline
DAC sampling rate & ${B}/{L}$ & ${B}/{L}$ & $B$ & $B$ & $B$        \\ \hline
Max. unamb. range & $r_\text{\textcolor{black}{max}}$& ${r_\text{\textcolor{black}{max}}}/{L}$ & \makecell{${r_\text{\textcolor{black}{max}} \cdot N_\text{\textcolor{black}{a}}}/{N_\text{\textcolor{black}{c}}}$ \\ ($<{r_\text{\textcolor{black}{max}}}/{L}$)} & $r_\text{\textcolor{black}{max}}$  & ${r_\text{\textcolor{black}{max}}}/{L_\text{\textcolor{black}{c}}}$     \\ \hline
Max. unamb. velocity & ${v_\text{\textcolor{black}{max}}}/{L}$ & $v_\text{\textcolor{black}{max}}$ & $v_\text{\textcolor{black}{max}}$ & $v_\text{\textcolor{black}{max}}$  & ${v_\text{\textcolor{black}{max}}}/{L_\text{\textcolor{black}{s}}}$       \\ \hline
Processing gain & ${N_\text{\textcolor{black}{c}} N_\text{\textcolor{black}{s}}}$ & ${N_\text{\textcolor{black}{c}} N_\text{\textcolor{black}{s}}}/{L}$ & \makecell{$N_\text{\textcolor{black}{a}} N_\text{\textcolor{black}{s}}$ \\ $(<{N_\text{\textcolor{black}{c}} N_\text{\textcolor{black}{s}}}/{L})$} & $N_\text{\textcolor{black}{c}} N_\text{\textcolor{black}{s}}$ & $N_\text{\textcolor{black}{c}} N_\text{\textcolor{black}{s}}$ \\ \hline
\textcolor{black}{Noise folding factor} & $\times$ & $\times$ & \textcolor{black}{$L$} & \textcolor{black}{$L$} & \textcolor{black}{$L$} \\ \hline
Additional hardware & Fast-settling PLL for LO stepping & Frequency comb generator & $\times$ & $\times$ & $\times$    \\ \hline
Additional processing & Phase offset calibration & Phase offset calibration & $\times$ & Symbol-mismatch noise cancellation & $\times$\\ 
\bottomrule
\end{tabular}
\begin{minipage}{17.4cm}
\vspace{0.1cm}
\small  $^{*}$ $N_\text{a}$ $(<{N_\text{c}}/{L})$ is the number of active subcarriers, determined as in \cite{SA_OFDM}.
\end{minipage}
\end{table*}

Fig. \ref{f8} presents a comparison of range profiles and Doppler profiles between the SNS-OFDM \textcolor{black}{\cite{SNS_OFDM}} and PC-SNS-OFDM radar systems. The ADC sampling rate, $L$, $L_\text{\textcolor{black}{c}}$, and $L_\text{\textcolor{black}{s}}$ remain consistent with those in Fig. \ref{f7}. The noise levels in both cases are identical to those in Fig. \ref{f7}. On the other hand, while the SNS-OFDM system exhibits peaks solely at the target's location, the PC-SNS-OFDM system shows the occurrence of three ambiguous peaks in both the range profiles and the Doppler profiles.

Fig. \ref{f9} demonstrates range and Doppler profiles obtained by varying the combinations of $L_\text{\textcolor{black}{c}}$ and $L_\text{\textcolor{black}{s}}$ in the proposed PC-SNS-OFDM system. \textcolor{black}{Ambiguous peaks do not appear in this figure. Only the maximum unambiguous range and velocity are depicted.} In the case where $L$ is set to 16, there are five possible combinations of $L_\text{\textcolor{black}{c}}$ and $L_\text{\textcolor{black}{s}}$ that satisfy the condition \textcolor{black}{$L = L_\text{\textcolor{black}{c}} \cdot L_\text{\textcolor{black}{s}} $}. As the values of $L_\text{\textcolor{black}{c}}$ and $L_\text{\textcolor{black}{s}}$ change, the maximum unambiguous range, $r_\text{u}$, becomes $r_\text{\textcolor{black}{max}}/L_\text{\textcolor{black}{c}}$, while the maximum unambiguous velocity becomes $v_\text{\textcolor{black}{max}}/L_\text{\textcolor{black}{s}}$. These graphs demonstrate the flexibility of adjusting the range and Doppler ambiguities within the PC-SNS-OFDM radar system. Consequently, the waveform can be modified adaptively based on target conditions to have the best measurements.

\subsection{Comparisons with other OFDM radars}
In this section, we present comparisons of the proposed method with other radar systems. 
\textcolor{black}{Since a method to reduce the sampling rate in other non-OFDM digital radars has not been explored yet, the comparison with the other non-OFDM digital radars at the same ADC sampling rate is not included here.}
The sparse OFDM radar system is excluded because it is not suitable as a low-cost radar system due to the large amount of computational effort when using CS. \textcolor{black}{
The proposed system has limitations to use in communication, since the use of the same symbol across all sub-bands results in the decrease in bit rate. Also, the noise folding due to sub-Nyquist sampling reduces the bit error rate (BER). Therefore, the following comparisons are only for a radar application.}
\subsubsection{Stepped-carrier (SC) OFDM}
SC-OFDM\textcolor{black}{\cite{SC_OFDM2}} is a radar system that employs a sequential transmission approach by dividing a wide-bandwidth OFDM signal into narrow-bandwidth sub-bands. This division allows for a reduction in the sampling rates of both the \textcolor{black}{digital-to-analog converter (DAC)} and the ADC. On the other hand, the PC-SNS-OFDM radar system can solely decrease the sampling rate of the ADC. In the SC-OFDM case, with each transmission of the sub-blocks, the frequency of the \textcolor{black}{LO} in the SC-OFDM system must change. Consequently, the inclusion of a fast-settling \textcolor{black}{PLL} becomes essential to control the LO frequency. Conversely, because PC-SNS-OFDM adopts the same hardware configuration used in a conventional OFDM radar system, there is no introduction of additional noise from the hardware. Moreover, due to its structural characteristic of dividing one OFDM symbol into several time slots for transmission, SC-OFDM results in a reduction in the maximum unambiguous velocity by a factor of $L$, where $L$ is the number of sub-blocks. 
\subsubsection{Frequency comb (FC) OFDM}
FC-OFDM\textcolor{black}{\cite{FC_OFDM}} is a radar system structure that enables an expansion of the bandwidth by performing up-conversion and down-conversion of a baseband OFDM signal with a narrow bandwidth into a combination of carrier frequencies. Unlike the PC-SNS-OFDM radar system, which only reduces the sampling rate of the ADC, FC-OFDM has the advantage of reducing both the DAC and ADC sampling frequencies. However, FC-OFDM entails a complex up-down conversion hardware structure due to the utilization of multiple carrier frequency combinations. In contrast, the PC-SNS-OFDM radar system features a simpler hardware configuration. Another characteristic of FC-OFDM is that only every \textcolor{black}{$L\text{th}$} subcarrier is utilized, with those remaining left as empty carriers. As a result, the maximum unambiguous range is reduced by a factor $L$. Additionally, during the down-conversion process, signals from a wide band overlap into the same band, causing the folding of the noise that existed within the same bandwidth. The increased noise level equals the folded noise level observed in the PC-SNS-OFDM radar system. FC-OFDM introduces another challenge in the form of phase offsets between each sub-band, as it employs multiple carrier frequency combinations. Consequently, a calibration process is required to correct these phase offsets. If the offsets are not accurately removed, the peak-to-sidelobe ratio increases considerably. With regard to the processing gain, it is reduced by a factor of $L$ compared to the PC-SNS-OFDM radar system.

\subsubsection{Subcarrier aliasing (SA) OFDM}
SA-OFDM\textcolor{black}{\cite{SA_OFDM}} is a radar system that utilizes only the \textcolor{black}{$\mu\text{th}$} subcarrier as an active subcarrier, while the remaining subcarriers are \textcolor{black}{zero} subcarriers. Uniform sub-Nyquist sampling is performed in the ADC, resulting in the active subcarriers being aliased to the position of the zero subcarriers. Because SA-OFDM also employs zero subcarriers, the result in this case is a reduction of the maximum unambiguous range. Here, the reduction ratio of the maximum unambiguous range is greater than the reduction ratio of the ADC sampling rate, as the active subcarriers should be precisely aliased to the position of the zero subcarriers and the interval between active subcarriers should be greater than the reduction rate of the ADC sampling rate. In contrast, the PC-SNS-OFDM radar system exhibits the same reduction ratio for both the ADC sampling rate and the maximum unambiguous range and velocity. Both the SA-OFDM and the PC-SNS-OFDM radar systems show the noise-folding effect. If the number of active subcarriers is denoted as $N_\text{\textcolor{black}{a}}$, the processing gain in SA-OFDM becomes $N_\text{\textcolor{black}{a}}N_\text{\textcolor{black}{s}}$, which is lower than that of the PC-SNS-OFDM radar system.

\subsubsection{Sub-Nyquist sampling (SNS) OFDM}
\textcolor{black}{The SNS-OFDM radar in \cite{SNS_OFDM} employs a conventional OFDM waveform, while the proposed PC-SNS-OFDM radar utilizes a phase-coded OFDM waveform for sub-band multiplexing. The SNS-OFDM radar introduces symbol-mismatch noises due to signal folding incurred by sub-sampling, which must be canceled out by additional iterative processing. The symbol-mismatch noise cancellation requires increased complexity for OFDM radar signal processing compared to that of the conventional one. Also, the number of iterations increases with the number of targets. On the other hand, the proposed PC-SNS-OFDM radar system takes advantage of the coded OFDM waveform, which can multiplex each sub-band. This allows the separation of each sub-band from the folded signal after sub-Nyquist sampling. Consequently, it does not require any additional processing to recover the sub-sampled OFDM signal at the receiver side, while scarifying the maximum unambiguous range or Doppler according to the selection of $L_{c}$ and $L_{s}$.  }

Table \ref{t2} provides a comparison of the radar systems discussed above. The systems were evaluated based on a standardized RF bandwidth, denoted here as $B$, while assuming a total of $N_\text{\textcolor{black}{c}}$ subcarriers and $N_\text{\textcolor{black}{s}}$ OFDM symbols. Furthermore, the ADC sampling rate reduction ratio is uniformly set to $L$ for all systems. Specifically, the SC-OFDM system employed $L$ sub-bands, the FC-OFDM system utilized a $L$ comb of frequencies, and the PC-SNS-OFDM system divided the RF bandwidth into $L$ sub-bands. The SA-OFDM system operated with $N_\text{\textcolor{black}{a}}$ active subcarriers. $r_\text{\textcolor{black}{max}}$ and $v_\text{\textcolor{black}{max}}$ represent the maximum range and velocity of a general OFDM radar system, respectively.

\begin{table}[t!]
\centering
\fontsize{8.3}{14}\selectfont
\caption{OFDM radar parameters for measurements}
\label{t3}
\begin{tabular}{>{\centering\arraybackslash} m{20em} | >{\centering\arraybackslash} m{7em}}
    \toprule
    Parameters    & Values\\ \hline
    \midrule
Number of subcarriers, $N_\text{\textcolor{black}{c}}$         & 2048         \\ \hline
Number of symbols, $N_\text{\textcolor{black}{s}}$             & 10           \\ \hline
Subcarrier spacing, $\Delta f$                  & 488 kHz       \\ \hline
Total signal bandwidth, B                       & 1 GHz     \\ \hline
OFDM symbol duration, $T_\text{\textcolor{black}{sym}}$          & 2.05 \textcolor{black}{$\upmu$}s    \\ \hline
CP duration,  $T_\text{\textcolor{black}{CP}}$                   & 0.51 \textcolor{black}{$\upmu$}s \\ \hline
OFDM symbol duration with CP, $T_\text{\textcolor{black}{s}}$    & 2.56 \textcolor{black}{$\upmu$}s \\ \hline
Carrier frequency, $f_\text{0}$               & 60.98 GHz  \\
    \bottomrule
\end{tabular}
\end{table}

\begin{figure}[t!]
    \centering
    \subfloat[]{\includegraphics[scale=0.43]{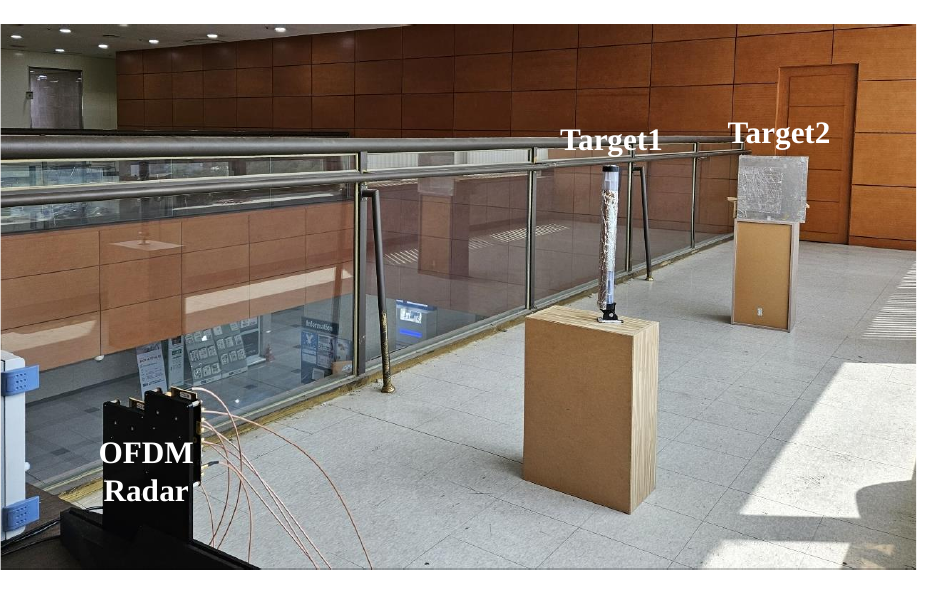}}

    \centering
    \subfloat[]{\includegraphics[scale=0.45]{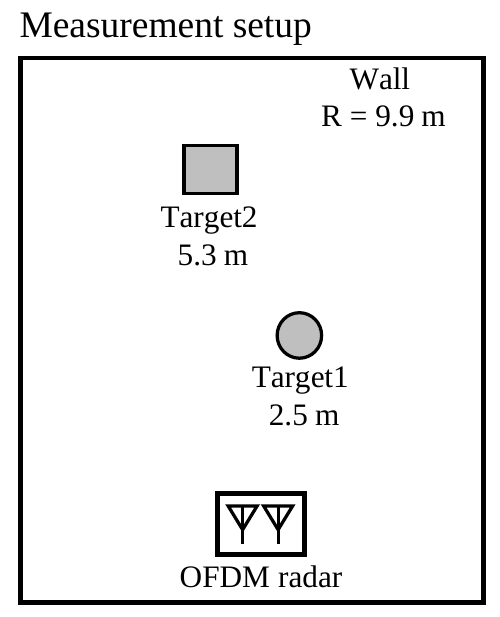}}
    
    \caption{(a) Photograph and (b) arrangement of the measurement setups}
    \label{f10}
\end{figure}

\begin{figure}[t!]
    \centering\includegraphics[scale=0.51, angle = 270]{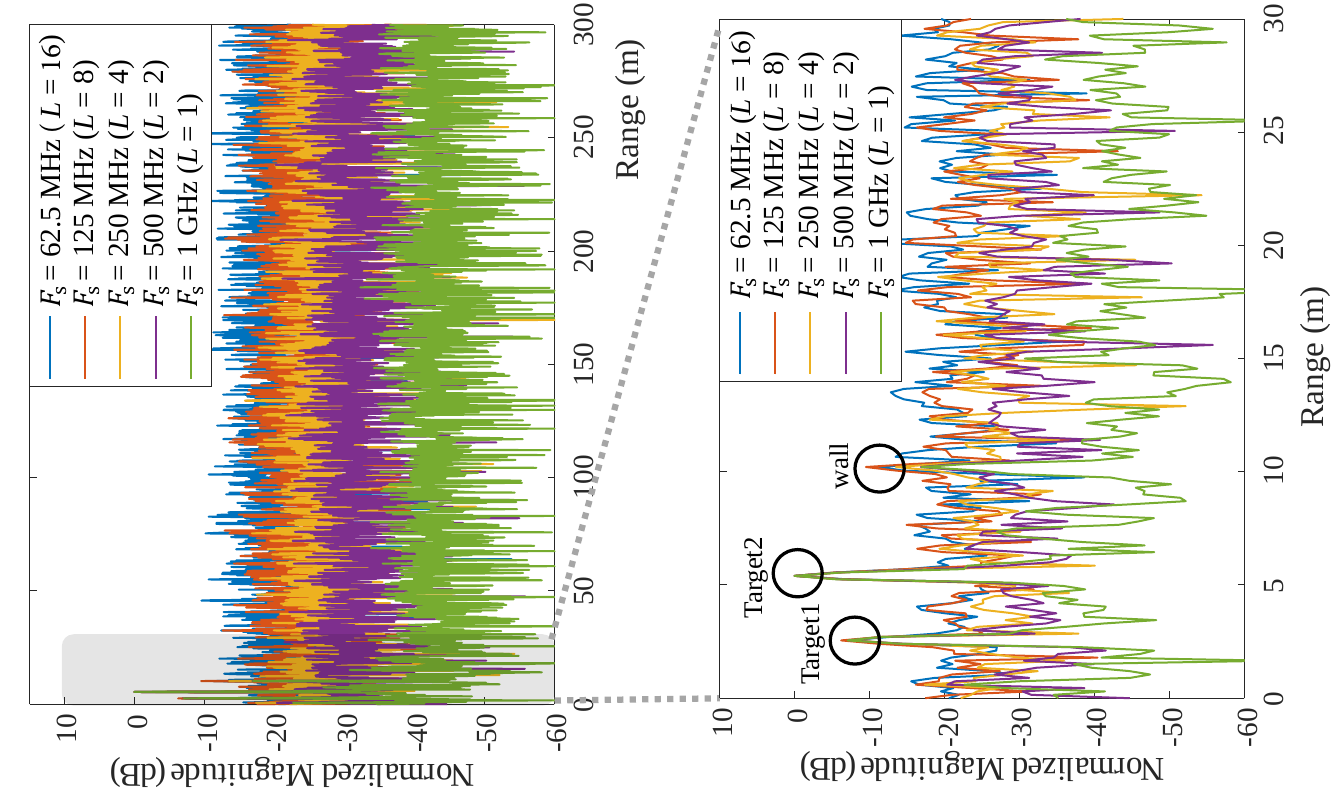}
    \caption{Range profiles of the \textcolor{black}{SNS-OFDM} radar system under various ADC sampling rates}
    \label{f11}
\end{figure}

\section{Measurements Results of PC-SNS-OFDM Radar System}
\textcolor{black}{In this section, we validate the proposed PC-SNS-OFDM radar system with various sub-sampling ratios and show the effectiveness of the phase-coded waveform in the sub-Nyquist sampling OFDM radar. Experimental comparisons with other OFDM radar systems having the reduced ADC sampling rate such as SC-OFDM and FC-OFDM radar are not investigated here, because the hardware configurations substantially differ from the conventional OFDM radar system.}

Table \ref{t3} shows the OFDM radar parameters for the measurements. The radar signals are composed of 2048 subcarriers and ten OFDM symbols. The bandwidth of the signal is 1 GHz, and the subcarrier spacing is the bandwidth divided by the number of subcarriers. The OFDM signals have a range resolution of 15 cm and a maximum detectable range of 307.2 m. Due to limitations of the experimental setups, it is not feasible to measure moving targets. As a result, the phase code of the time domain is excluded from these measurements. Specifically, $L_\text{\textcolor{black}{s}}$ is set to 1, while only the value of $L_\text{\textcolor{black}{c}}$ varies with changes in the sampling rate of the ADC.

\begin{figure}[h!]
    \centering\includegraphics[scale=0.53, angle = 270]{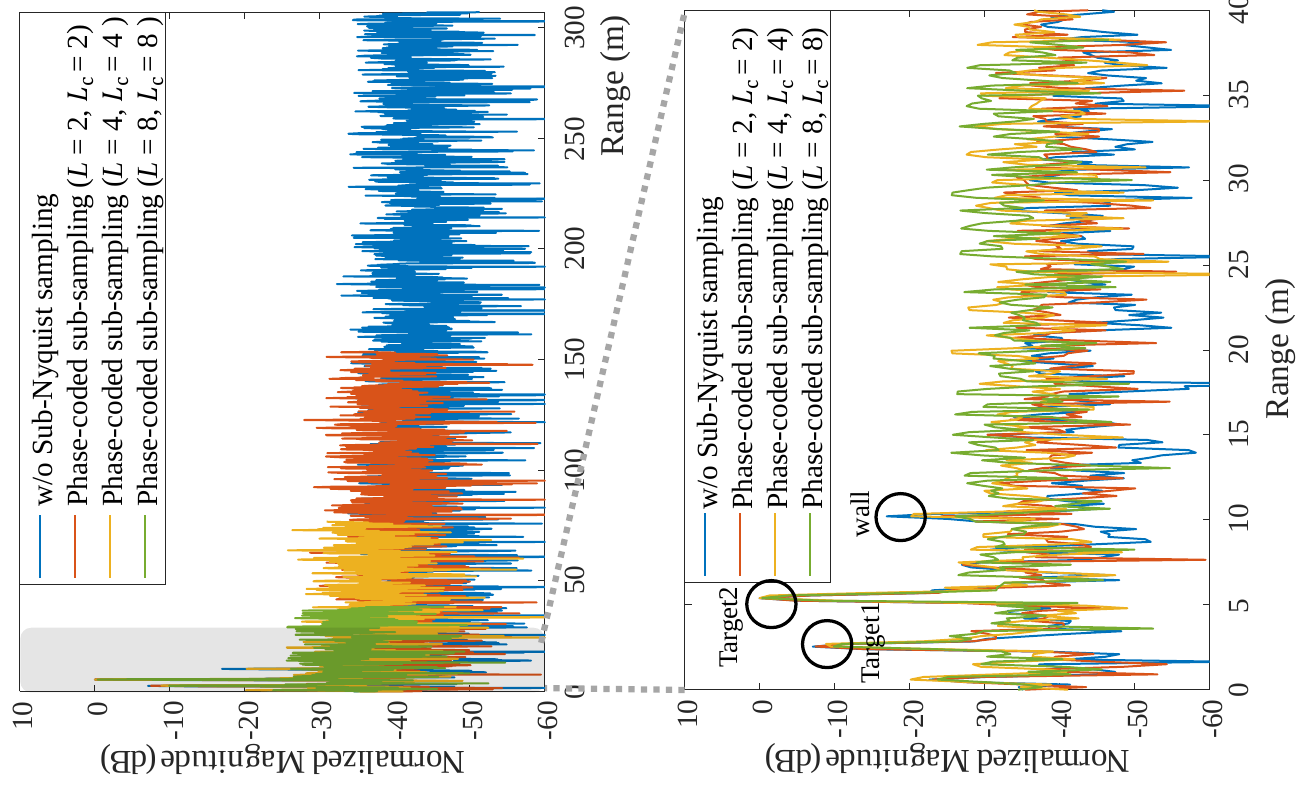}
    \caption{Range profiles of the \textcolor{black}{PC-SNS-OFDM} radar system under various ADC sampling rates}
    \label{f12}
\end{figure}

Fig. \ref{f10} illustrates the measurement setups employed in this paper, established within an indoor environment. Two targets are positioned at distances of 2.5 m and 5.3 m from the radar, while a wall is located at a distance of 9.9 m from the radar. A commercial beamforming module with a 60 GHz carrier frequency is utilized as the OFDM radar. The DAC employed is the M8190A arbitrary waveform generator (AWG), which operates at a sampling rate of 8 GHz. The baseband signal generated by the AWG is up-converted to the carrier frequency at the transmit radar module. Subsequently, the signal reflected from the targets is down-converted to a baseband signal at the receive radar module. This baseband signal is then transmitted to an oscilloscope, where it is sampled at various sampling frequencies and converted to a digital signal. With the digital signal, signal unfolding and radar processing are conducted.

\begin{figure}[t!]
    \centering
    \subfloat[]{\includegraphics[scale=0.65]{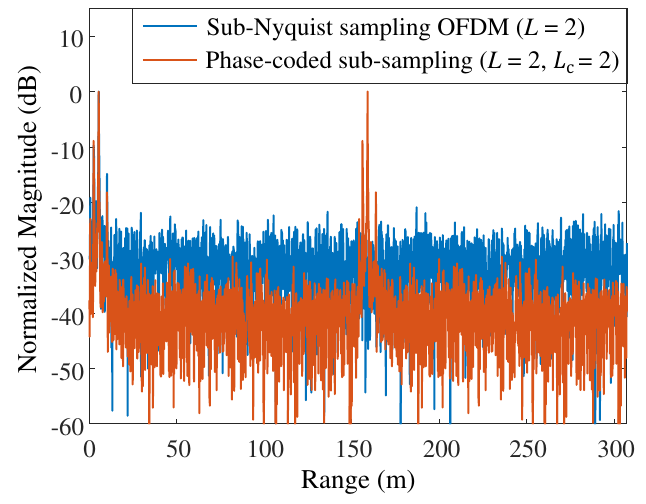}}
    
    \centering
    \subfloat[]{\includegraphics[scale=0.65]{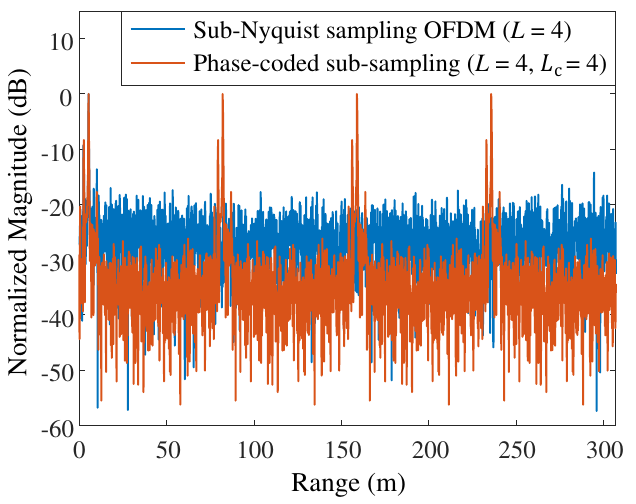}}

    \centering
    \subfloat[]{\includegraphics[scale=0.65]{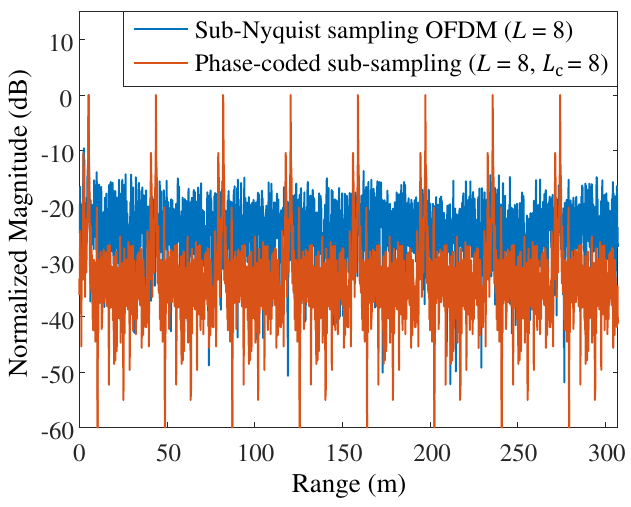}}

    \caption{Range profiles of the SNS-OFDM and the proposed PC-SNS-OFDM radar systems under various $L$: (a) \textcolor{black}{$L = 2$, (b) $L = 4$, and (c) $L = 8$}}
    \label{f13}
\end{figure}

\begin{figure}[t!]
    \centering
    \subfloat[]{\includegraphics[scale=0.65]{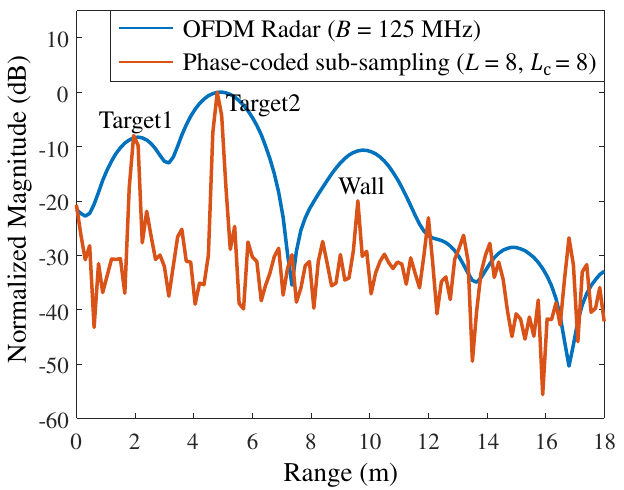}}
    
    \centering
    \subfloat[]{\includegraphics[scale=0.65]{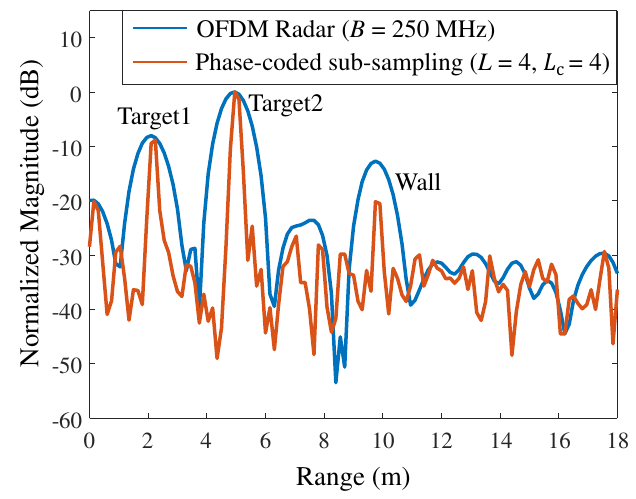}}

    \centering
    \subfloat[]{\includegraphics[scale=0.65]{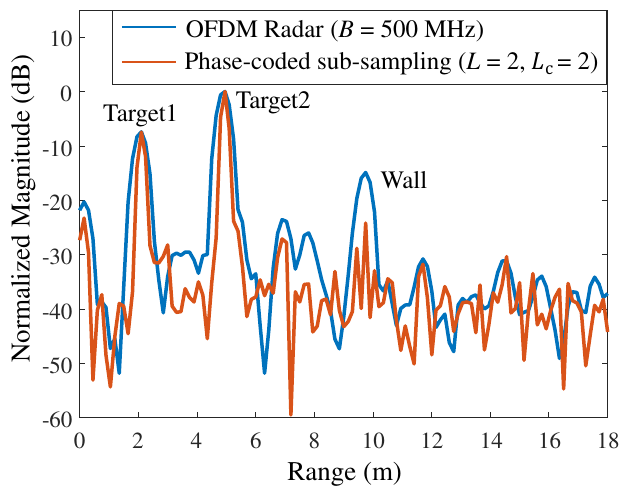}}

    \caption{Range resolutions of the PC-SNS-OFDM radar and the conventional OFDM radar without sub-Nyquist sampling at the same ADC sampling rate: (a) 125 MHz, (b) 250 MHz, and (c) 500 MHz}
    \label{f14}
\end{figure}

The measurement results of the SNS-OFDM system \textcolor{black}{\cite{SNS_OFDM}}, with respect to various ADC sampling rates, are presented in Fig. \ref{f11}. The range resolution of the SNS-OFDM system remains equivalent to that of an OFDM signal with a 1 GHz bandwidth. Note that the noise level at $L = 1$, equivalent to a general OFDM radar system, is measured at \textcolor{black}{$-44.8$} dB, serving as a reference. However, as the ADC sampling rate decreases, the noise level increases. To be specific, these outcomes are \textcolor{black}{$-32.8$} dB for $L = 2$, \textcolor{black}{$-27.8$} dB for $L = 4$, \textcolor{black}{$-24.5$} dB for $L = 8$, and \textcolor{black}{$-22.1$} dB for $L = 16$. The experimental results demonstrate that a decrease in the sampling rate of the ADC leads to an increase in the number of folded signals, resulting in higher noise levels.

The measurement results obtained from the PC-SNS OFDM system are presented in Fig. \ref{f12}. Note that the PC-SNS-OFDM radar system demonstrates a consistent range resolution equivalent to that of an OFDM signal with a 1 GHz bandwidth, irrespective of variations in the ADC sampling rate. A decrease in the ADC sampling rate results in an increase in the noise level due to noise folding. Specifically, the system exhibits noise levels of \textcolor{black}{$-41.5$} dB for $L = 2$, \textcolor{black}{$-35.4$} dB for $L = 4$, and \textcolor{black}{$-33.4$} dB for $L = 8$. However, it is essential to note that symbol-mismatch noise does not occur, suggesting that the increase in the noise level is not significant. Instead, due to range ambiguity, the maximum unambiguous range decreases to $R_\text{\textcolor{black}{max}}/L_\text{\textcolor{black}{c}}$.

Fig. \ref{f13} presents the results of a comparative analysis of the measurement outcomes obtained from the SNS-OFDM \cite{SNS_OFDM} and PC-SNS-OFDM systems at the same ADC sampling rate. In the first graph, the noise level is \textcolor{black}{$-32.8$} dB for SNS-OFDM and \textcolor{black}{$-41.5$} dB for PC-SNS-OFDM. Moving to the second graph, the noise level for SNS-OFDM is \textcolor{black}{$-27.8$} dB, while PC-SNS-OFDM records a noise level of \textcolor{black}{$-35.4$} dB. Finally, in the last graph, SNS-OFDM exhibits a noise level of \textcolor{black}{$-24.5$} dB, while the PC-SNS-OFDM system shows a noise level of \textcolor{black}{$-33.4$} dB. The proposed method displays a noise level approximately 8 to 9 dB lower than that of the SNS-OFDM radar system. However, as the ADC sampling rate decreases, there is an increase in the number of ambiguous peaks. Specifically, $L-1$ ambiguity peaks occur per target. 

Fig. \ref{f14} presents graphs comparing the range resolution between a conventional OFDM radar system and the PC-SNS-OFDM radar system, again with the same ADC sampling rate. In this comparison, the bandwidth of the OFDM radar is set to $F_\text{\textcolor{black}{s}}$, the maximum bandwidth achievable based on Nyquist sampling theory. It is observed that in the conventional OFDM radar system, a decrease in the ADC sampling rate necessitates a reduction in the signal bandwidth, degrading the range resolution. However, with the proposed PC-SNS-OFDM radar system, a high range resolution can still be achieved, even with a lower ADC sampling rate.

\section{Conclusions}
This paper presents a PC-SNS-OFDM radar system designed to achieve a high range resolution while utilizing low ADC sampling rates. A time-frequency phase-coded OFDM waveform is proposed, which effectively eliminates symbol-mismatch noise in SNS-OFDM without requiring a complex symbol-mismatch cancellation process. Furthermore, a flexible waveform generation method is introduced, enabling adaptive control of range and Doppler ambiguities depending on the measured target situation. Comparisons with alternative methods that also attempt to reduce the ADC sampling rate in OFDM radar are conducted, revealing that the proposed system has advantages over other reported approaches in terms of hardware and signal processing. In addition, the proposed method is the only technology capable of providing adaptive control of range and Doppler ambiguities. 

\ifCLASSOPTIONcaptionsoff
  \newpage
\fi



\bibliographystyle{IEEEtran}
\bibliography{IEEEabrv,reference}
%



%





\vfill


\end{document}